\documentclass[11pt,a4paper]{article}
\usepackage{authblk}
\pdfoutput=1
\usepackage{subcaption}
\usepackage{graphicx}
\usepackage{pdfpages}
\usepackage{setspace}
\usepackage{geometry}
\usepackage{graphicx}
\usepackage{float}
\usepackage{amsmath}
\usepackage{amssymb}
\usepackage{hyperref}
\usepackage{titlesec}
\usepackage{color}
\usepackage{tikzpagenodes}
\usepackage{tikz}
\usepackage{caption}
\usepackage{threeparttable}
\usepackage{background}
\usepackage[numbers,sort&compress,comma]{natbib}
\definecolor{chaptercolor}{RGB}{0,0,0}             
\definecolor{sectioncolor}{RGB}{0,0,0}          
\definecolor{subsectioncolor}{RGB}{0,0,0}           

\titleformat{\title}[hang]
  {\normalfont\bfseries\color{chaptercolor}\fontsize{22}{24}\selectfont}
  {}
  {0pt}
  {\MakeUppercase}
\titlespacing{\title}{0pt}{12pt}{12pt}

\titleformat{\chapter}[hang]
  {\normalfont\bfseries\color{chaptercolor}\fontsize{16}{18}\selectfont}
  {\thechapter.}
  {0.5em}
  {}
\titlespacing{\chapter}{0pt}{12pt}{12pt}

\titleformat{\section}[hang]
  {\normalfont\bfseries\color{sectioncolor}\fontsize{14}{16}\selectfont}
  {\thesection}
  {0.5em}
  {}
\titlespacing{\section}{0pt}{12pt}{12pt}

\titleformat{\subsection}[hang]
  {\normalfont\color{subsectioncolor}\fontsize{12}{14}\selectfont}
  {\thesubsection}
  {0.5em}
  {}
\titlespacing{\subsection}{0pt}{12pt}{12pt}

\backgroundsetup{
   scale=1,
   angle=0,
   opacity=1,
   color=black,
   contents={\begin{tikzpicture}[remember picture,overlay]
        \node[anchor=north east, inner sep=0, align=right] at ([xshift=-2.5cm, yshift=-1cm]current page.north east) {}; 
        \node[anchor=north east, inner sep=0, align=right] at ([xshift=-2.5cm, yshift=-1.5cm]current page.north east) {}; 
    \end{tikzpicture}
   \\\vspace{1cm} }
}

\begin{document}



# Wavefront Shaping of Scattering Forces Enhances Optical Trapping of Levitated Nanoparticles

Melissa Kleine[1], Michael Horodynski[2], Stefan Rotter[3], Yacine Amarouchene[1], Yann Louyer[1], Mathias Perrin[1], Nicolas Bachelard[1]*

[1]CNRS, Université de Bordeaux, LOMA, UMR5798, Talence, France.

[2]Department of Physics, Massachusetts Institute of Technology, Cambridge, MA, 02139, USA

[3]Institute for Theoretical Physics, Vienna University of Technology (TU Wien), A–1040 Vienna, Austria

*email: nicolas.bachelard@u-bordeaux.fr

Abstract: Optically-levitated nanoparticles in vacuum offer a pristine platform for high-quality mechanical oscillators, enabling a wide range of precision measurements and quantum technologies. A key performance metric in such systems is the stiffness of the optical trap, which is typically enhanced by increasing laser power—at the cost of unwanted heating, reduced coherence, and enhanced quantum backaction. Here, we demonstrate a fundamentally new route to increasing trap stiffness: wavefront shaping of the optical field. By tailoring the spatial phase profile of the trapping beam, we significantly boost the mechanical confinement of subwavelength particles without raising the optical intensity. Remarkably, this enhancement arises from a selective reduction of non-conservative optical forces, while preserving the conservative restoring forces that define trap stiffness. As a result, mechanical nonlinearities are also reduced, improving stability at low pressures. Our findings challenge the long-standing assumption that diffraction-limited focusing is optimal for dipolar Rayleigh particles, and establish wavefront shaping as a powerful, readily applicable tool to control optomechanical forces in levitation experiments. This opens new avenues for minimizing backaction, reducing thermal decoherence, and expanding the range of materials that can be stably levitated.

Optical levitation offers the tantalizing prospect of harnessing quantum effects at the mesoscale for both fundamental as well as technological purposes, and multiple milestones have recently been reached in the pursuit of these goals [1]. Through the focusing of a laser beam, particles with sizes in the nanometer or micrometer range can be trapped by an optical tweezer in high vacuum, resulting in underdamped oscillations around an equilibrium position. In contrast to other optomechanical devices [2], the stiffness of the oscillator that governs its dynamics, scales here linearly with laser intensity, providing remarkable tunability. Large stiffnesses are, in particular, required to overcome decoherence rates and perform ground-state cooling of nanoparticles [3–6] as well as to achieve high-performance sensors [7]. Importantly, trapping has to be photon-efficient—i.e., provide maximal stiffness for minimal optical intensity, as levitated objects in high vacuum are known to heat up substantially due to residual absorption, which is detrimental for quantum coherence [8] and can lead to particle melting [9]. Operating at reduced power is also of major importance to mitigate photonic recoil that hampers cooling [10]. Several theoretical proposals have therefore been put forward to increase a trap's stiffness at fixed or reduced intensity, for example by using different materials and/or particles' dimensions [11,12]. Yet, no experimental realization of these concepts with optically levitated nano-objects in high vacuum has been reported so far.

In practice, the optomechanical stiffness is governed by the interplay between the different optical forces produced by the trapping field, which divide into conservative (gradient) and nonconservative (scattering) components [13]. Gradient forces attract the particle towards intensity maxima, whereas scattering forces commonly push it away along the optical axis, thus weakening optical confinement. For microscale particles (i.e., dimensions largely exceeding the wavelength), which exhibit numerous Mie resonances [14], stiffness is strongly hampered by significant scattering forces, whose contributions can be easily tailored—e.g., to balance gravity [15]. In contrast, levitated subwavelength nanoparticles are often approximated as simple Rayleigh scatterers (or point-like dipoles) with negligible mass, for which gradient forces dominate and lead to increased stiffnesses. Nevertheless, due to an inherently complex dipolar polarizability, the scattering force remains nonzero, shifting the equilibrium position away from focus [16], where field intensity—and therefore stiffness—is reduced. It has long been assumed that, unlike for microparticles, the relative contributions of gradient and scattering forces in a diffraction-limited laser field provide optimal confinement of nanoparticles, thus leaving no straightforward route for improvement.

In domains like multiple-scattering media [17–19] or complex beam generation [20], wavefront-shaping approaches, relying on devices such as Spatial Light Modulators (SLM), have been tailored to control optical-field distributions. In particular, in the case of single microparticles, a spatially-modulated wavefront can reorganize the contributions of Mie resonances, thus reshaping light-matter interactions [21]. Notably, for large microparticles trapped in liquid—where motion is overdamped—spatially modulating the trapping field's phase has successfully been used to enhance optomechanical stiffness [22,23]. However, this strategy is commonly believed to be ineffective for nanoparticles levitated in vacuum, which, unlike microparticles trapped in water, are expected to behave as point-like Rayleigh scatterers, displaying underdamped motion on very short timescales. Given these constraints, it remains unclear whether wavefront shaping can provide any meaningful control over these particles' optomechanical confinement.

In this work, we challenge and overturn this prevailing assumption by demonstrating that optomechanical stiffness in nanoparticle levitation can indeed be enhanced through wavefront shaping. We achieve this by spatially modulating the trapping beam's wavefront to reduce

2

scattering forces while preserving gradient forces. Specifically, we implement an iterative optimization routine that shapes the incoming beam onto a focusing objective to maximize the stiffness of a subwavelength silica bead, thereby making the optical tweezer more photon-efficient. Numerical simulations confirm that this optimization reshapes spatially the scattering force, shifting the equilibrium position closer to focus, where the particle experiences stronger confinement. Experimentally, we validate this effect by analyzing modifications within the stochastic toroidal motions embedded into the particle's dynamics, revealing direct signatures of reduced nonconservative forces. Crucially, our findings demonstrate that acting on scattering forces—previously considered a fixed constraint—provides a viable pathway to enhance optical confinement even at the nanoscale. Importantly, as optomechanical confinement is enhanced and the particle moves closer to focus, we also observe a reduction in motional nonlinearities that influence the overdamped dynamics at low pressures.

We experimentally implement a wavefront-shaping approach that restructures the focused trapping field in order to increase the stiffness of deeply-subwavelength levitated particles. As displayed in Figure 1**a**, we levitate individual nanoparticles (radius $R$ ranging from 75 to 125 $nm$, red dot) inside a vacuum chamber (**Supplementary Figure S1**) using an objective lens ($NA = 0.8$), which focuses an incoming laser (wavelength $\lambda = 1064\ nm$, intensity $\approx 300\ mW$, red beam) whose phase is modulated through a phase-only SLM (**Methods**). The SLM encodes non-uniform wavefronts, $\phi(r)$, which reshape the focused field while preserving its overall laser intensity. Below a pressure of a few mbar, the particle experiences underdamped oscillations along $x$, $y$ and $z$, which materialize as mechanical resonances in the Power Spectral Densities (PSDs), $S_{jj, j \in [x,y,z]}$, emerging for a uniform (i.e., unmodulated) wavefront at around 100, 120 and 30 kHz for $x$, $y$ and $z$, respectively (**Supplementary Figure S2**). The corresponding stiffnesses $\kappa_x$, $\kappa_y$ and $\kappa_z$ are readily deduced from their respective resonance frequencies. The black curve in Figure 1**b** displays the axial PSD, $S_{zz}$, of a 125 $nm$-radius silica particle levitated using a uniform wavefront (black circle, inset). The beam's phase is then iteratively shaped through an optimization routine and converges towards an "optimized wavefront" (white to black concentric circles, inset), which simultaneously improves the axial frequency resonance (along $z$, red curve) and the transverse ones (along $x$ and $y$, **Supplementary Figure S3**). The corresponding iterative enhancements in optical confinements performed throughout the optimization are reported in Figure 1**c**, where, at each iteration, the ratios of the modulated stiffness $\kappa_{opt, i \in [x,y,z]}$ over their initial (i.e., uniform) values $\kappa_{0, i \in [x,y,z]}$ are displayed to converge towards final magnifications of 1.4, 1.5 and 2.5 along $x$, $y$ and $z$, respectively. When applied to nanoparticles of different sizes, our approach systematically enhances the stiffness (colored curves, Figure 1**d**), requiring different optimized wavefronts (white to black insets). While the transverse stiffness enhancements follow similar trends when the size increases (yellow and green curves), the axial enhancement turns out to be systematically larger (blue curve). Interestingly, larger particles—bestowed with larger scattering cross sections—display stronger stiffness enhancements along all directions. This emphasizes that the optimization is more efficient on objects that are subject to stronger scattering forces. Remarkably, the wavefronts are customized to match the particle they have been optimized for and appear only partially efficient when applied to particles of different sizes (**Supplementary Figure S4**). Our optimization is implemented through a gradient-free routine that maximizes a cost function assembled from the stiffness ratios, while the wavefronts are expanded over a Zernike-polynomial basis (**Supplementary Figure S5**). The cost function can be adapted to reach different ratios and promote specific resonance directions (**Supplementary Figure S6**). This routine proves very robust: Even though it can converge towards different final solutions associated with different optimized wavefronts, we constantly report comparable enhancements for a given cost function.

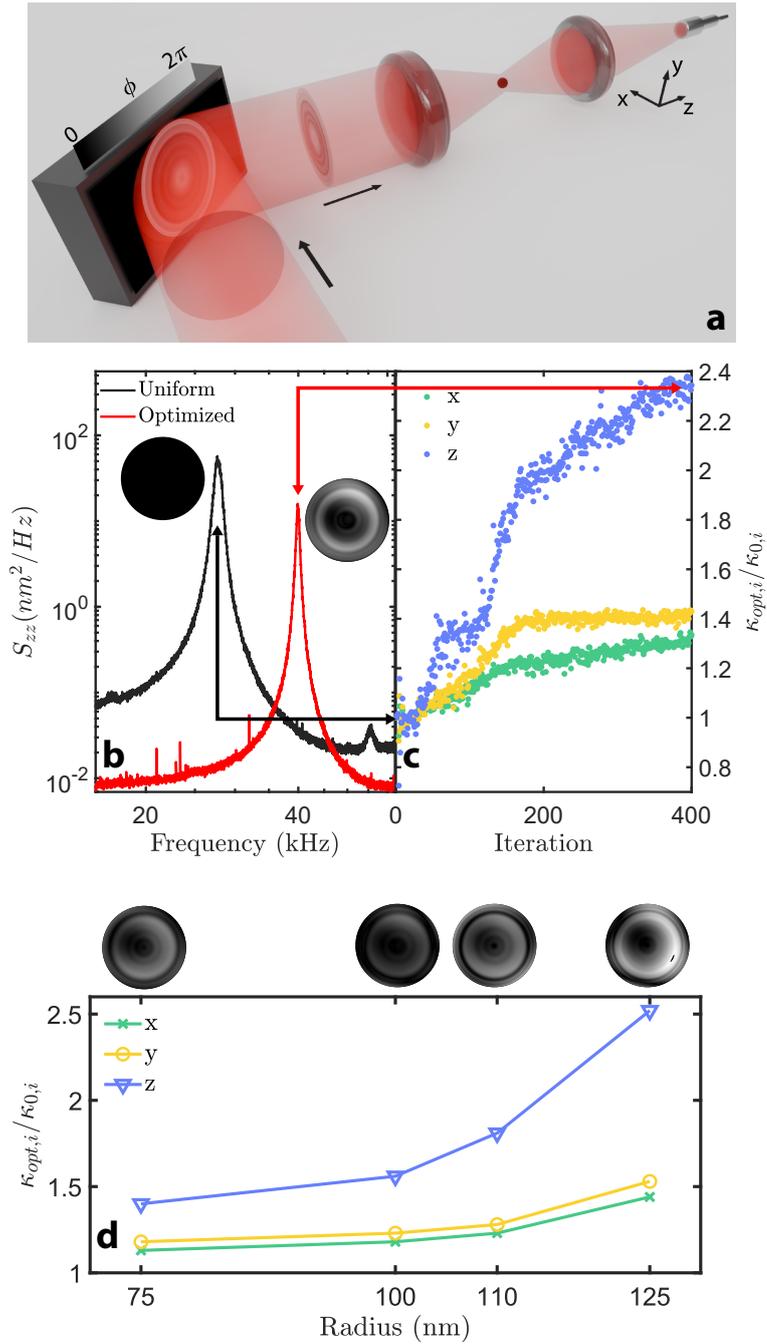

Figure 1 : **Experimental optimization**. **a**, A laser (red beam) is focused by a high-NA objective to levitate nanoparticles (red dot) in vacuum. An SLM (black box) phase-modulates the beam and imprints a non-uniform wavefront (white to black concentric circles, $\phi(r)$ with $r$ the spatial coordinate on the SLM) that reshapes the trapping field (i.e., focused field). The transmitted light is collected by a second lens and injected into a fiber bundle to read out the nanoparticle's motion along $x$, $y$ and $z$. **b**, PSDs $S_{zz}$ of a $125\ nm$-radius nanoparticle, measured for a uniform (black curve, black circle in inset) and an optimized wavefront (red curve, black to white circles in inset). **c**, Evolution of the stiffness ratios $\kappa_{opt,x}/\kappa_{0,x}$ (green), $\kappa_{opt,y}/\kappa_{0,y}$ (yellow) and $\kappa_{opt,z}/\kappa_{0,z}$ (blue) throughout the optimization procedure. The iterations used in **b** are marked by the black and red double-arrow lines. **d**, Optimized stiffness ratios obtained after applying the routine to nanoparticles of different radii (same color code as in **c**). The corresponding optimized wavefronts are reported in the insets.

Simulations emphasize that, along the optical axis, the optimization effectively brings the equilibrium position closer to the focus where optical intensity is larger and optomechanical confinement is enhanced. The optimization performed in Figure 1**b** and **c** is emulated through a numerical model, which computes, for arbitrary wavefronts, the exact optical forces' landscapes through an accurate expansion onto electric and magnetic multipoles [14] (**Supplementary Figure S7** and **Methods**). The black curves on Figure 2**a** and **b** display respectively the total axial force, $F_Z$, exerted onto a 125 $nm$-radius particle by a uniform and an optimized wavefront (insets), when its position is varied along the optical axis, $z$. Equilibrium locations ($z_{eq}$, vertical grey dashed line) coincide with the positions where the force cancels ($F_Z(z_{eq}) = 0$) and around which the particle oscillates following a dynamic governed by the trap stiffness $\kappa_Z = -d_Z F_Z(z_{eq})$ (grey dotted-dashed line). As expected, for a uniform wavefront, the scattering force pushes the nanoparticle away from focus to an equilibrium position $z_{eq} \approx 2.0\ \mu m$. We observe that the optimized wavefront reshapes the axial force's landscape, which brings the particle closer to focus at $z_{eq} \approx 1.6\ \mu m$, where the stiffness is increased by roughly a factor of 2.2 with respect to the uniform case. As expected, the trap is also more photon-efficient as the optimized wavefront can produce the same value of $\kappa_z$ as in the uniform case but with now half the intensity (**Supplementary Figure S8**). Once the particle lies closer to focus, we also improve the stiffnesses $\kappa_x$ and $\kappa_y$ along the transverse directions $x$ and $y$, respectively (**Supplementary Figure S9**). Our model also highlights that, even if the nanoparticle behaves as a Rayleigh scatterer (Mie parameter $2\pi R/\lambda \approx 0.74$), electric and magnetic multipoles do play a role. The forces $F_z$ displayed in Figure 2**a** and **b** are expanded into their main multipolar contributions, corresponding to electric-dipole ($F_{z,ED}$, blue), magnetic-dipole ($F_{z,MD}$, red) and electric-quadrupole ($F_{z,EQ}$, orange) terms. Yet, while we observe that higher terms are required to stabilize trapping in the uniform case of Figure 2**a**, they have minor contributions to the mechanical stiffness that is mainly supported by the electric dipole (i.e., larger slope). As a result, the optimized wavefront mainly reshapes the electric-dipole landscape to improve the optomechanical stiffness $\kappa_z$, leaving higher terms unaffected (Figure 2**b**).

Our numerical model also emphasizes that, as sketched in Figure 2**c**, the optimized wavefront improves the stiffness by preserving the conservative gradient force while simultaneously strongly reshaping the non-conservative scattering force's landscape. Along the optical axis, a uniform wavefront (black inset) generates an almost Gaussian-like trapping potential ($U_0$, black line), in which scattering forces lead to the nanoparticle (gray dot) being trapped at a location $z_{eq}$ away from focus where confinement softens. The optimized wavefront (white to black inset) barely modifies the potential ($U_{opt}$, red), while reducing the scattering. Thus, the nanoparticle is brought closer to focus, at a location where the potential quickly stiffens. This mechanism is illustrated in Figure 2**d**, where the total optical forces $F_Z$ computed in Figure 2**a** and Figure 2**b** are decomposed into their gradient ($F_{g,0} = -d_z U_0$ and $F_{g,opt} = -d_z U_{opt}$, dashed lines) and scattering components ($F_{s,0}$ and $F_{s,opt}$, solid lines), displayed in black and red for a uniform and an optimized wavefront, respectively. We observe that the gradient-force landscape (i.e., potential's derivative) remains largely unaffected by the optimization, while the scattering force is strongly shifted towards focus. This enforces a new equilibrium position, $z_{eq}$, where the slope of the gradient force significantly contributes to increase the total stiffness. The estimation of the gradient and scattering force landscapes is described and discussed in detail in **Supplementary Figure S10**.

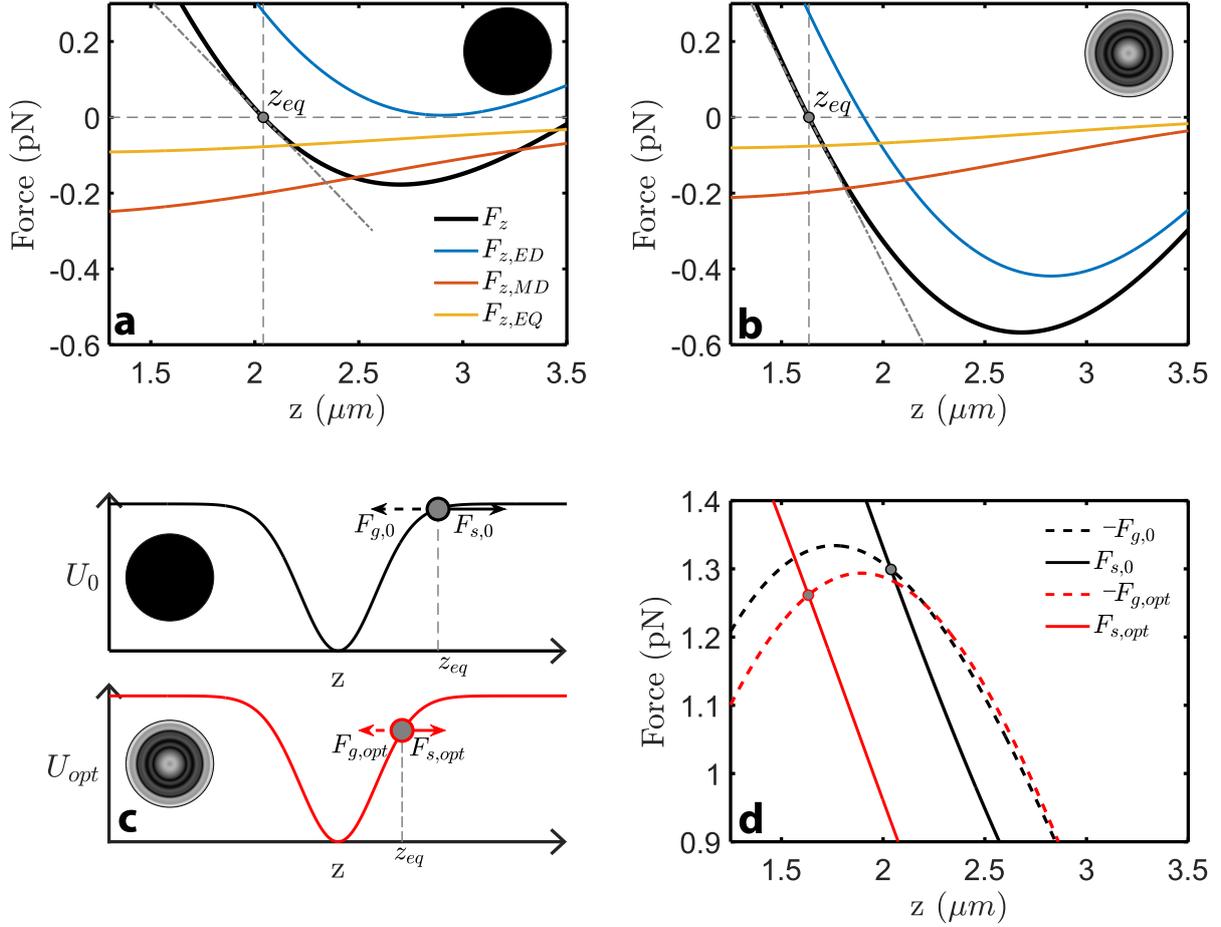

Figure 2: **Numerical simulations of force landscapes**. **a,b**, Computed axial-force landscape ($F_z$, black) for a uniform and an optimized wavefront (insets) that are focused onto a 125 $nm$-radius nanoparticle. The forces are expanded into an electric-dipole ($F_{z,ED}$, blue), a magnetic-dipole ($F_{z,MD}$, red) and an electric-quadrupole ($F_{z,EQ}$, orange) contributions, which include interference terms (**Supplementary Figure S7**) and fulfill $F_z = F_{z,ED} + F_{z,MD} + F_{z,EQ}$. Equilibrium positions (vertical grey dashed) fulfill $F_z(z_{eq}) = 0$, while the stiffness (grey dashed-dotted) defines as $\kappa_z = -d_z F_z(z_{eq})$. **c**, Along the optical axis, $z$, a uniform beam (black circle) produces an almost-gaussian optical potential ($U_0$, black) that sets the gradient force landscape ($F_{g,0} = -d_z U_0$, dashed black arrow). The nanoparticle (grey bead) is trapped at an equilibrium position ($z_{eq}$) shifted away from focus, where $F_{g,0}$ balances the scattering force ($F_{s,0}$, solid black arrow). The optimized wavefront (black to white concentric circles) barely modifies the potential ($U_{opt}$, red), while reducing the scattering force ($F_{s,opt}$, solid red arrow). The equilibrium is shifted to a location where the gradient force ($F_{g,opt} = -d_z U_{opt}$, dashed red arrow) matches $F_{s,opt}$ and the potential is significantly stiffer. **d**, The axial-force distributions, $F_z$, displayed in **a** and **b** are decomposed into the sum of their respective gradient ($F_{g,0}$, $F_{g,opt}$), and scattering parts ($F_{s,0}$, $F_{s,opt}$), fulfilling $F_z = F_g + F_s$. The black dashed and solid curves represent respectively $-F_{g,0}$ and $F_{s,0}$ that are obtained under a uniform wavefront and which intersect at the equilibrium position $z_{eq} = 2\ \mu m$. The red dashed and solid curves represent respectively $-F_{g,opt}$ and $F_{s,opt}$ that are obtained under an optimized wavefront and which intersect at the equilibrium position $z_{eq} = 1.6\ \mu m$.

A stochastic analysis of the nanoparticle's non-equilibrium dynamics confirms experimentally that wavefront shaping achieves a significant modulation of the scattering forces. Due to their non-conservative nature, scattering forces generate toroidal stochastic motions, whose extension reveals much smaller than the particle size [16,24–27]. Yet, these motions materialize, both on the position and velocity probability distributions, as probability currents describing Brownian vortices (**Supplementary Figure S11**). Denoting by $v_\rho$ and $v_z$ the norms of the transverse and axial velocities of a $125\ nm$-radius nanoparticle, Figure 3**a** displays a histogram of the probability distribution ($P_v$, colored map) together with its current ($\bar{J}_v(v_\rho, v_z)$, black arrows), which are assembled from a $10\ s$-long trajectory obtained under a uniform wavefront (inset). Figure 3**b** reproduces the same analysis with the optimized wavefront of Figure 1**b** (inset) and shows, as expected, confinement enhancements along both the transverse and axial directions, which are combined with a strong current (i.e., vortex) reduction indicative of a significant scattering-force adjustment. In Figure 3**c**, the variance, $\langle \bar{J}_v^2 \rangle$, of the currents displayed in panels **a** and **b** are monitored when the pressure is varied, which emphasizes that this reduction remains when the dynamics of the particle varies. Yet, while current modifications are indicative of an adjustment in non-conservative forces, their interpretation remains qualitative (**Supplementary Figure S12**).

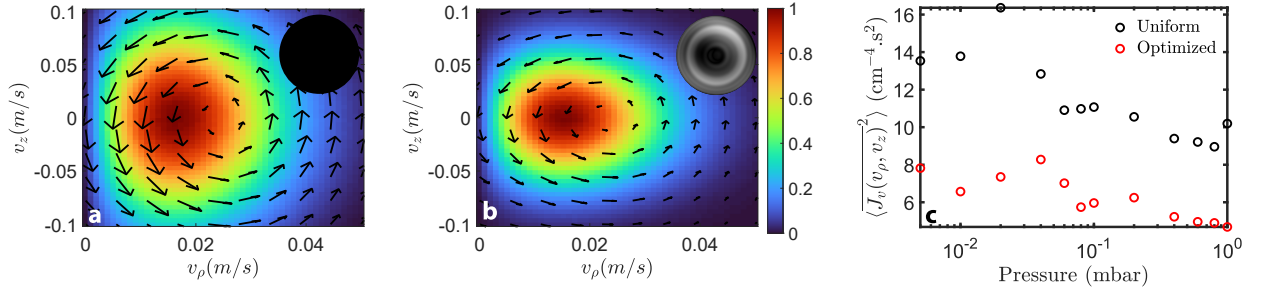

Figure 3 : **Experimental observation of scattering-force modulation**. **a,b**, Probability distributions $P_v$ (normalized by their maximal values, colored map) and their respective currents $\bar{J}_v$ (black arrows) at 1 mbar in the velocity plane $(v_\rho, v_z)$ for the uniform and optimized wavefronts of Figure 1**b**, respectively (insets). **c**, Averaged flux variance, $\langle \bar{J}_v^2 \rangle$, as a function of pressure for the uniform (black) and optimized wavefronts (red) of **a** and **b**, respectively.

At last, we observe experimentally that improving the stiffness also delays the emergence of Duffing non-linearities responsible for spectral deformations at low pressures [24,28]. When motional damping drops, thermalized particles start to explore the non-linear regions of the trapping landscape. Therefore, stronger confinements (constraining the oscillations' amplitude) and trapping the particle closer to the focus (i.e., further away from non-linearities) effectively reduce the influence of non-linearities. As Duffing spectral deformations are more pronounced in the transverse plane, we plot in Figure 4**a** the PSDs $S_{yy}$ for different pressures, measured for a $125\ nm$-radius nanoparticle and while applying the uniform (black) and optimized wavefronts (red) of Figure 1 and Figure 3. At a pressure of 10 mbar, the two PSDs are almost-Lorentzian, while below 0.1 mbar the one obtained under an optimized wavefront is clearly less subject to non-linearities. Interestingly, more quantitative insights on the reduction of non-linearities can be gained through short-term dynamics. For instance, at 0.1 mbar and under a uniform wavefront, Figure 4**b** provides $S_{yy}$ assembled from 5 ms and 20 s trajectories (orange/yellow dashed and black, respectively, which emphasizes that spectral broadening and the subsequent non-linear deformations originate from stochastic shifts of the short-time resonance frequency. In Figure 4**c**, when pressure is varied, we plot the standard deviations of the short-time-resonance distributions along $x$, $y$ and $z$, which we assemble from frequency-distribution histograms (**Supplementary Figure S13**). As expected, when applying the optimized wavefront, non-linearities—which are characterized by large deviations in short-time-resonance—are reduced below roughly 1 mbar (grey regions).

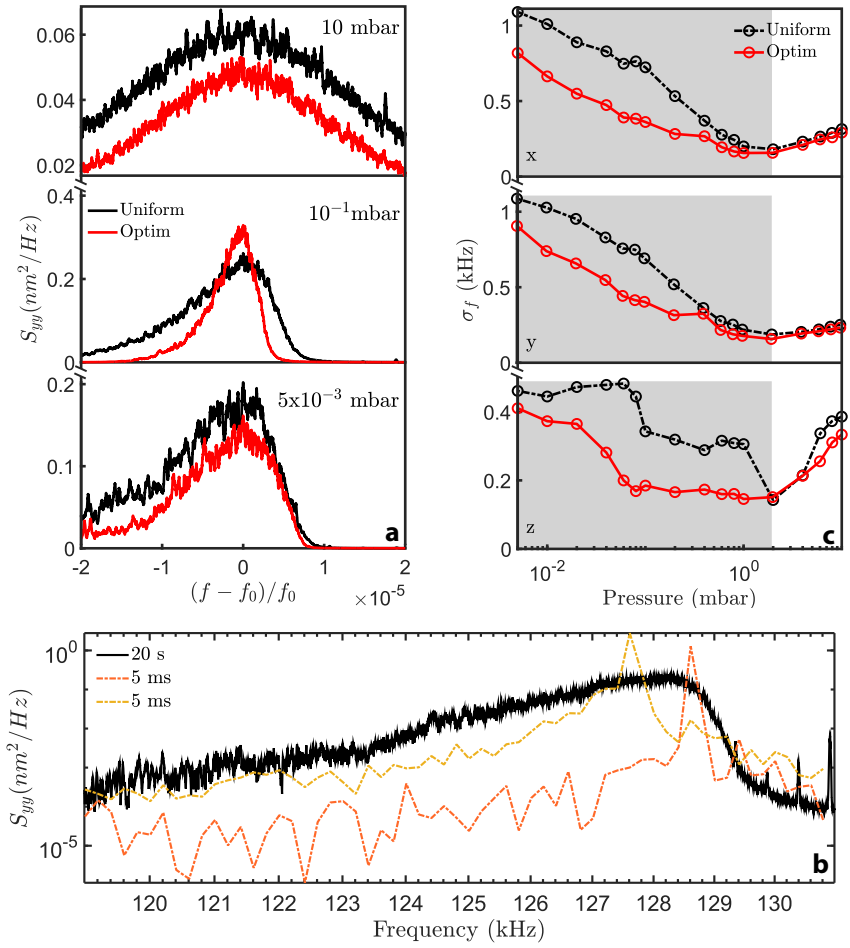

Figure 4 : **Experimental reduction of non-linearities**. **a**, PSDs $S_{yy}$ of a 125 $nm$-radius sphere measured at three different pressures under the uniform (black) and optimized (red) wavefronts of Figure 1 and Figure 3. The reduction of spectral broadening in the optimized case for pressures below 1 $mbar$ emphasizes that nonlinearities are strongly reduced. **b**, PSDs $S_{yy}$ obtained for a 20 $s$-long time trace (black) and two Fourier-limited 5 ms-long time traces (dotted red, dotted orange). **c**, Standard deviations, $\sigma_f$, of 5 $ms$ time-trace frequency distributions measured along $x$, $y$ and $z$ at different pressures. The black and red curves are obtained using the wavefronts of **a** (same color code). The grey regions indicate where $\sigma_f$ is lower (and nonlinearities reduced) for the optimized wavefront compared to the uniform case.

In conclusion, we have demonstrated that wavefront shaping can enhance the mechanical confinement of levitated nanoparticles by modulating non-conservative scattering forces. By leveraging commercially available spatial light modulators (SLMs), our approach can be seamlessly integrated into existing optical levitation setups, enabling more photon-efficient confinement. This allows for high trapping stiffness with minimal optical power, a crucial advantage in levitation experiments where photon efficiency directly impacts particle heating and backaction. While wavefront shaping has previously been explored for levitating single microparticles [29] or multiple microparticles [30,31], our findings underscore its potential for trapping single nanoparticles—key to advancing quantum applications [1]. Building on the remarkable progress in wavefront shaping [32,33], we anticipate that this technique will help address major challenges in optical levitation, including stable trapping in high vacuum and the levitation of particles composed of diverse materials [34], complex geometries [35] or embedded with emitters [36,37].

Methods

Experimental setup:

Inside a vacuum chamber of about $5\,L$, the light produced by a linearly polarized, ultra-low-noise and water-cooled laser (Azurlight System, @ $1064\,nm$, $10\,W$) overfills the back aperture of an Olympus LMPlan IRx100 objective ($NA = 0.8$) with a filling factor of $f_0 = 1.1$. This forms an optical tweezer with a beam waist of 430 nm, in which silica nanoparticles (Microparticles GMBH) are trapped. The laser beam is modulated using a phase-only SLM (Holoeye, Pluto-2.1 NIR 149), whose zeroth diffraction order is spatially filtered using a 4f system. The light scattered from the particle is collected using a $NA = 0.55$ aspheric lens and injected onto a four-quadrant-like photodetector made of a 1 to 4 multimode fiber bundle (Thorlabs BF42HS01) coupled to two balanced photodiodes (Thorlabs PDB440C-AC). The signal is digitized using a 350 MHz-bandwidth DAQ (GaGe, TB3-RazorMax, 500 MS/s, 16 bits). The optimization is performed using a Nelder-Mead simplex algorithm, which corresponds to a gradient-free routine. The wavefronts are expanded along a basis made of 30 Zernike polynomials. The cost function is assembled from the ratios of the square of resonance frequencies obtained when applying the wavefront and in the absence of any modulation.

Numerical model:

We compute the optical field focused onto the nanoparticle using a modified version of the Debye's integral [13], in which the wavefront produced by the SLM is modelled through a thin-lens apodization function describing a non-uniform phase distribution. The scattered field is obtained using the Generalized Lorentz Mie Theory (GLMT), as an expansion on vectorial spherical harmonics. Integrating the Maxwell Stress Tensor over a spherical shell surrounding the nanoparticle, we obtain the exact expression of the total optical force landscape [38]. In parallel, the total force is also obtained through a computation of the electric and magnetic multipoles up to quadrupole contributions [39,40], which matches GLMT calculations with great precision (**Supplementary Figure S7**). This multipolar approach gives analytical expressions of the force that reveal to be extremely useful to compute the conservative and non-conservative contribution [41] as well as to speed up the optical-force estimations required to emulate the experimental optimization procedure.



\title{Supplementary Information: Wavefront Shaping of Scattering Forces Enhances Optical Trapping of Levitated Nanoparticles}
\author[1]{Melissa Kleine}
\author[2]{Michael Horodynski}
\author[3]{Stefan Rotter}
\author[1]{Yacine Amarouchene}
\author[1]{Yann Louyer}
\author[1]{Mathias Perrin}
\author[1]{Nicolas Bachelard}
\affil[1]{Bordeaux University, CNRS, LOMA, UMR 5798, F-33405 Talence, France}
\affil[2]{Department of Physics, Massachusetts Institute of Technology, Cambridge, MA, 02139, USA}
\affil[3]{Institute for Theoretical Physics, Vienna University of Technology (TU Wien), Vienna, Austria}
\maketitle

\tableofcontents
\newpage
\section{Experimental setup}
\subsection{Optical trap}
\begin{figure}[h!]
    \centering
    \includegraphics[width=1\linewidth, trim={0cm 0cm 0cm 0cm},clip]{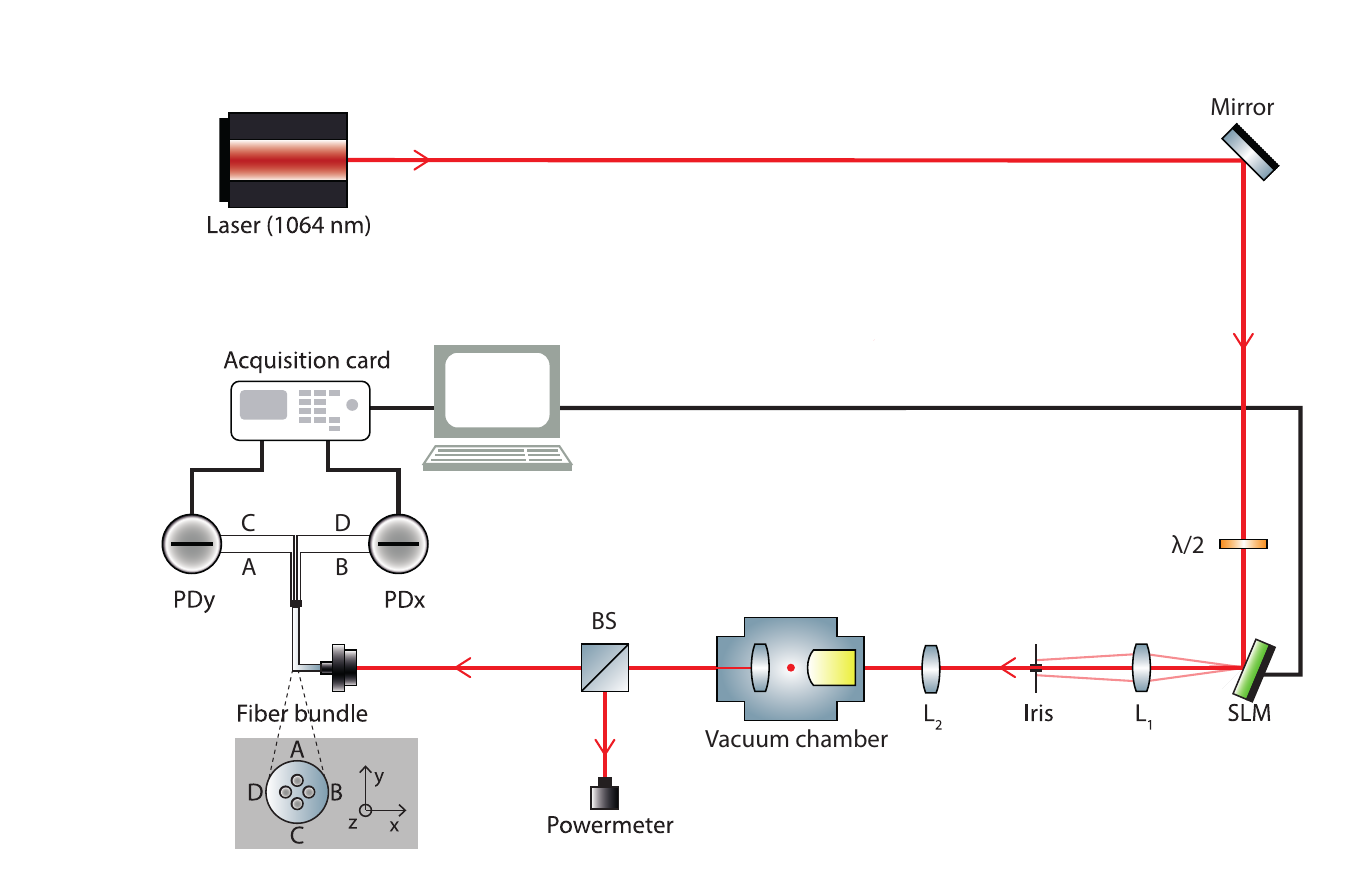}
    \caption{\textbf{Schematic representation of the experimental setup}. 
    The optical phase of a 1064 nm continuous laser beam is spatially shaped by an spatial light modulator (SLM) before being focused into a vacuum chamber using a high-numerical-aperture objective, thus forming an optical trap. 
    The beam is then recollected and analyzed using a differential photodiode system to extract motional power spectral densities (PSDs).}
    \label{fig:setup}
\end{figure}

Figure \ref{fig:setup} provides a schematic description of the experimental setup. 
A linearly polarized 1064 nm continuous laser (AzurLight System, 10 W) delivers roughly $\sim$300 mW of power at the input of a vacuum chamber. 
Inside the chamber, the beam is focused using a high-numerical-aperture objective (Olympus LMPlan IRx100, NA = 0.8, WD = 3.4 mm), forming a single-beam gradient optical trap, while an aspheric lens (NA = 0.55) collects the transmitted light.
The particles used in the experiment consist of silica nanobeads with radii of 75, 100, 110, and 125 nm (density: $\rho$ = 2200 kg.m$^{-3}$, refractive index: n = 1.45), sourced from \textit{Microparticles GmbH} and \textit{NanoCym}. 
Before trapping, they undergo a preheating process at 600°C for 3 hours. 
This treatment stabilizes the particles by eliminating Si-OH surface groups and forming durable Si-O-Si bonds\cite{art:chauffe_part}. 
A suspension of these particles in isopropanol is then sprayed into the chamber at atmospheric pressure using an \textit{Omron Micro-Air} nebulizer.

\begin{figure}[h!]
    \centering
    \includegraphics[width=1\linewidth, trim={0cm 0cm 0cm 0cm},clip]{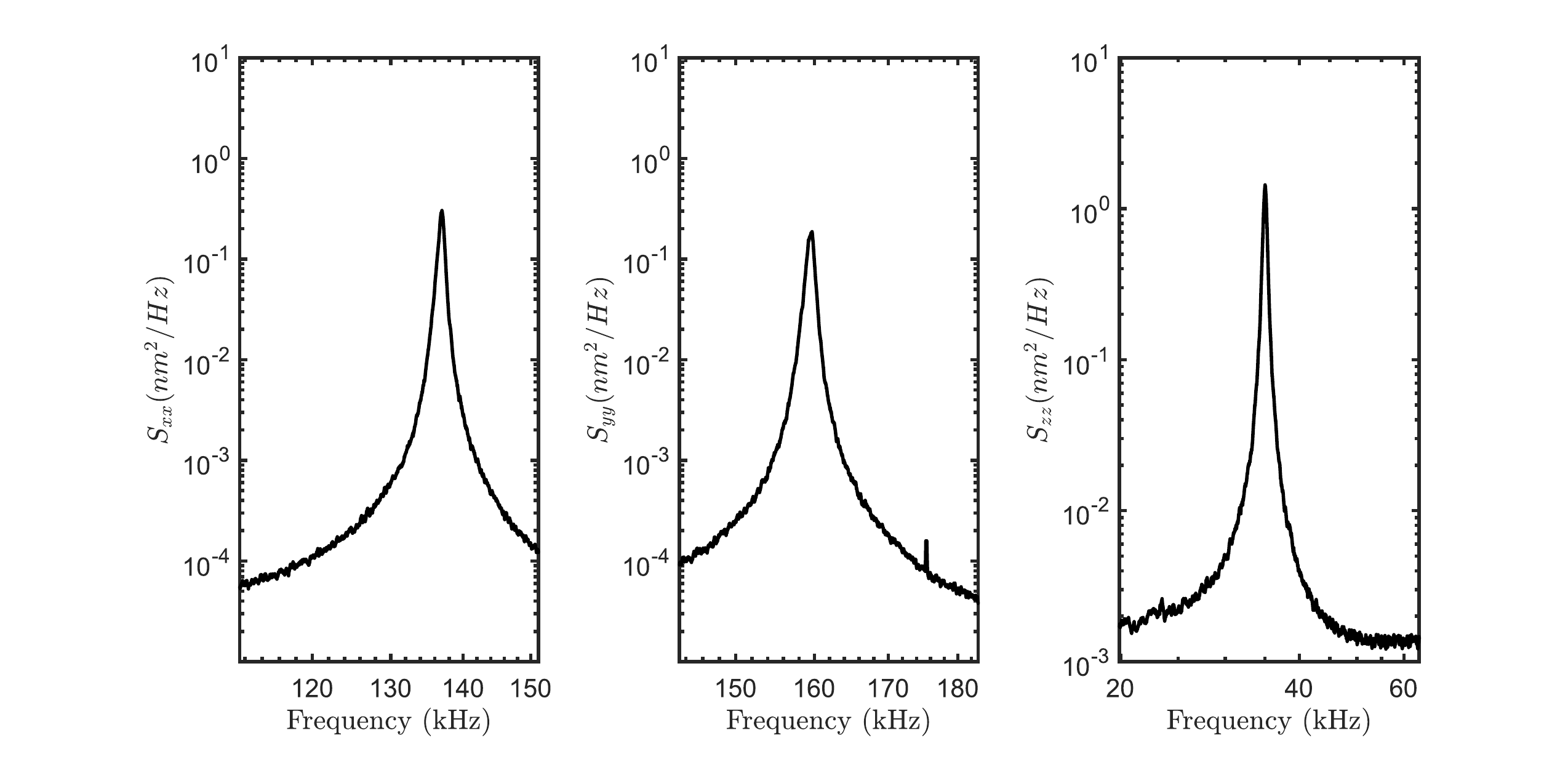}
    \caption{\textbf{Mechanical spectra measured in the absence of SLM modulation}. PSDs along all three axes for a 125 nm particle, at 1 mbar, trapped using a uniform wavefront and an optical power of $\approx 300 mW$ at the input of the vacuum chamber.}
    \label{fig:PSD_initial}
\end{figure}

The center-of-mass (COM) motion of the trapped particle is detected via spatial integration of the interference pattern formed between the trapping and scattered fields. 
As illustrated in Fig. \ref{fig:setup}, a split detection scheme, sensitive to transverse motion, is implemented by spatially dividing the beam using a 1-to-4 multimode fiber bundle, with each fiber being coupled to a differential photodiode. 
Motion along the optical axis can be measured on both photodiodes. 
COM displacements are recorded simultaneously along all three directions using a DAQ operating at a sampling rate of 5 MS/s and acquiring 20-second long time traces. 
Figure \ref{fig:PSD_initial} provides an example of measured power spectral densities (PSDs) along all three axes under a uniform wavefront (i.e., no modulation on the SLM).
\\

\subsection{Wavefront shaping}
We use a phase-only Spatial Light Modulator (SLM, Holoeye PLUTO-2.1 NIR149) with a resolution of 1920×1080 pixels and an 8 µm pixel pitch. 
The SLM modulates the phase of the beam before it passes through the trapping objective. 
Due to the Fourier transform relationship between the field at the SLM plane and the field at the focal plane of the objective, the phase-only modulation applied to the beam upstream translates (directly) into intensity modulations at the trap's location.
To fully exploit the capabilities of the SLM, it is initially configured into a blazed diffraction grating.
If this grating splits the beam into multiple diffraction orders, it reflects about $\approx 60 \% $ of the incoming  light intensity into the first order. 
Additional phase modulation patterns (of lower spatial frequencies) are then superimposed onto this blaze grating. 
According to Fourier optics, this superposition in the phase domain corresponds to a convolution of the patterns after the trapping objective, allowing the nonzero diffraction orders to be modulated. 
An iris is used to isolate the first diffraction order, which carries the desired phase-modulated beam, while eliminating the zero-order as well as higher-order contributions.

To determine the position of the incident beam on the SLM, we use a masking technique combined with the diffraction grating pattern.
The blaze grating pattern is applied only within a circular aperture; elsewhere, the phase modulation is set to zero (constant phase).
The radius of this circular mask is chosen to be smaller than the size of the incident beam.
The center of the circular mask is moved across the SLM surface in steps of 4 pixels.
For each position of the mask, we measure the power of the first diffraction order after the trapping objective.
The recorded power indicates the overlap between the incident beam and the modulated region of the SLM.
By scanning the mask position, we pinpoint the spatial location of the beam's center on the SLM.

\subsection{Stiffness optimization}
\label{sec:Opt}
The phase patterns or wavefronts used for stiffness optimization are constructed as linear combinations of Zernike polynomials.
As illustrated in Fig. \ref{fig:zernike_poly}, we use a basis made of 30 polynomials, whose first 20 elements are axis-symmetric ($Z_i^0$) while the remaining 10 are not ($Z_i^{k\ne 0}$). 
Zernike polynomials are particularly well-suited for wavefront shaping due to their orthogonality and their ability to accurately represent a wide range of optical aberrations.

\setcounter{figure}{4}
\begin{figure}[h!]
    \centering
    \includegraphics[scale=0.35]{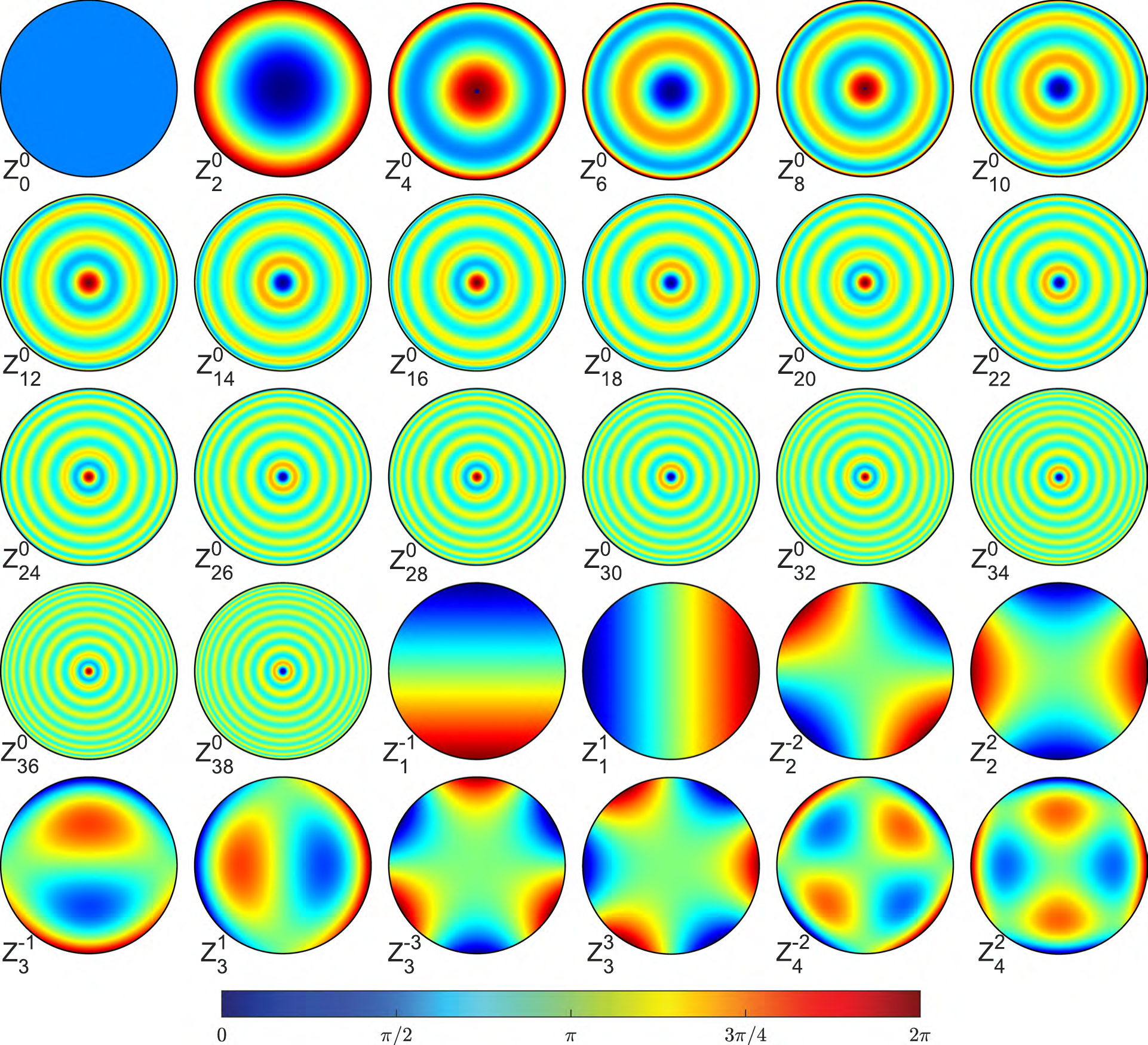}
    \caption{\textbf{Wavefront-shaping expansion basis}. {Zernike polynomials} used for the optimization protocol.}
    \label{fig:zernike_poly}
\end{figure}

In order to optimize the stiffness, we employ a gradient-free algorithm (\textit{fminsearch}, Matlab). 
This iterative algorithm adjusts the contribution of each Zernike polynomial to maximize a given cost function without requiring gradient calculations, which are known to be extremely sensitive to experimental noise. 
By avoiding abrupt phase changes, this approach also ensures smooth wavefront adjustments, thus reducing the risk of particle loss throughout the process.
For each iteration, an initial acquisition is performed using the uniform (i.e., unmodulated) beam to measure the baseline resonance frequencies, followed by a second acquisition with the applied phase pattern. 
Here, we recall that the resonance frequency $f_{i\in \{x,y,z\}}$ along each axis is related to the trap stiffness $\kappa_{i\in \{x,y,z\}}$ and the particle's mass $m$ through the relationship
\begin{equation}
    2\pi f_i=\sqrt{\frac{\kappa_i}{m}}
\end{equation}
We then estimate a cost function of the form 
\begin{equation}
    f(X) = \alpha(f_{opt,x}/f_{0,x})^2+\beta(f_{opt,y}/f_{0,y})^2+ \gamma(f_{opt,z}/f_{0,z})^2
    \label{eq:cost}
\end{equation}
, where $f_{0,i}$ and $f_{opt,i}$ represent respectively the resonance frequencies of the uniform and optimized traps, while the terms $\alpha$, $\beta$ and $\gamma$ stand for adjustable coefficients and $X$ denotes the wavefront (i.e., linear combination of Zernike polynomials used to generate the wavefront). 
The coefficients $\alpha$, $\beta$ and $\gamma$ can be adjusted to improve the optimization of one direction over the others.
This cost function is motivated by the proportionality between stiffness and the square of the resonance frequency as well as the fact that measuring the uniform resonance frequency at each iteration mitigates the impact of power drifts (thus ensuring an accurate evaluation of relative stiffness improvements).

\subsection{Experimental results}
Figure \ref{fig:iter} illustrates the evolution of the relative stiffness during three optimization routines performed on a particle of 110 nm in radius using three different cost functions. 
\setcounter{figure}{5}
\begin{figure}[h!]
    \centering
    \includegraphics[scale=0.4]{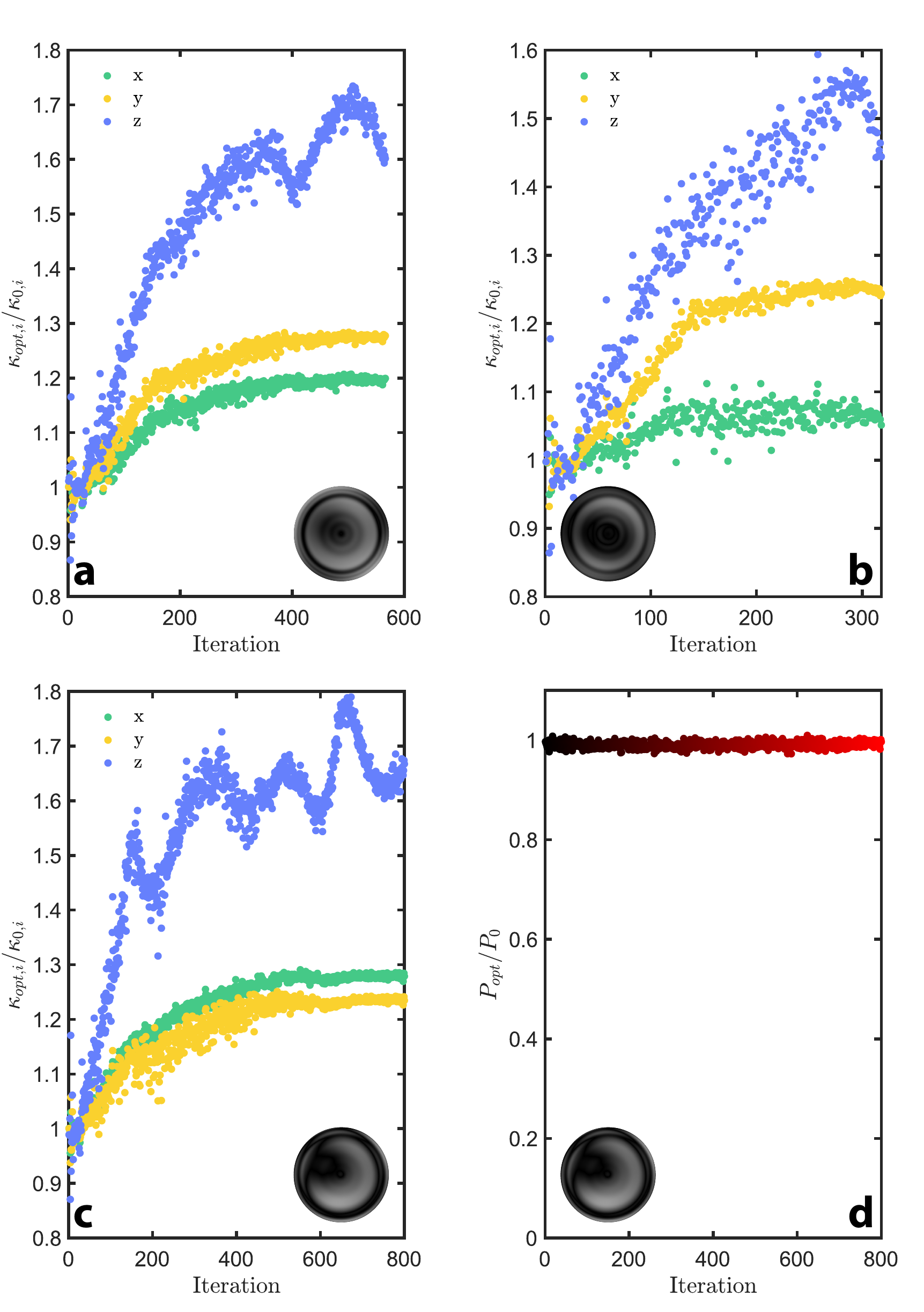}
    \caption{\textbf{Different examples of experimental optimizations}. 
    \textbf{a, b, c}- Evolution of the relative stiffness along each axis (see legend and color scheme) throughout the optimization for three different cost functions. 
    \textbf{d}- Evolution of the optical power, $P_{opt}$, at each iteration throughout the optimization displayed in \textbf{c}. 
    The power is measured by a photodiode positioned after the vacuum chamber and normalized by the value collected for a uniform wavefront, $P_0$.}
    \label{fig:iter}
\end{figure}

The first cost function (Fig. \ref{fig:iter}\textbf{a}), optimizes the relative stiffness along all three directions and reads
\begin{equation*}
    f(X)= (f_{opt,x}/f_{0,x})^2+(f_{opt,y}/f_{0,y})^2+\frac{1}{2}(f_{opt,z}/f_{0,z})^2
\end{equation*}
This results in relative stiffness ratios $\kappa_{opt,x}/\kappa_{0,x}=1.19$, $\kappa_{opt,y}/\kappa_{0,y}=1.26$ and $\kappa_{opt,z}/\kappa_{0,z}=1.73$.
Here, the coefficient $\gamma=1/2$ is introduced to mitigate optimization difficulties due to measurement uncertainties along $z$.
This cost function corresponds to the one used in Figure 1 of the main text in the case of 125-nm levitated particles. 

The second optimization (Fig. \ref{fig:iter}\textbf{b}) focuses solely on maximizing the stiffness along the $y$-axis, using the cost function
\begin{equation*}
    f(X)= (f_{opt,y}/f_{0,y})^2
\end{equation*}
While this also enhances stiffness along $z$, it has a negligible effect on $x$ and the corresponding stiffness ratios are respectively $\kappa_{opt,x}/\kappa_{0,x}=1.08$, $\kappa_{opt,y}/\kappa_{0,y}=1.26$ and $\kappa_{opt,z}/\kappa_{0,z}=1.57$.

Finally, the third optimization (Fig. \ref{fig:iter}\textbf{c}) targets the $x$-axis specifically, using the cost function-
  \begin{equation*}
      f(X)= (f_{opt,x}/f_{0,x})^2
  \end{equation*}
This approach leads to the highest relative stiffness along $x$, making it the only case where $\kappa_x>\kappa_y$, yielding ratios $\kappa_{opt,x}/\kappa_{0,x}=1.29$, $\kappa_{opt,y}/\kappa_{0,y}=1.24$ and $\kappa_{opt,z}/\kappa_{0,z}=1.79$.

The three resulting wavefronts, shown in Fig. \ref{fig:iter}, exhibit distinct differences, demonstrating the impact of the choice of the cost function on the optimized wavefront. 
Furthermore, Fig. \ref{fig:iter}\textbf{d} shows the evolution of the relative power measured by a photodiode located after the vacuum chamber (see Fig. \ref{fig:setup}) throughout the third optimization routine. 
The constant power level confirms that the optimization process does not affect the alignment of the beam or the filling factor of the trapping objective.

\setcounter{figure}{2}
\begin{figure}[h!]
    \centering
    \includegraphics[scale=0.325, trim={0cm 0cm 0cm 0cm},clip]{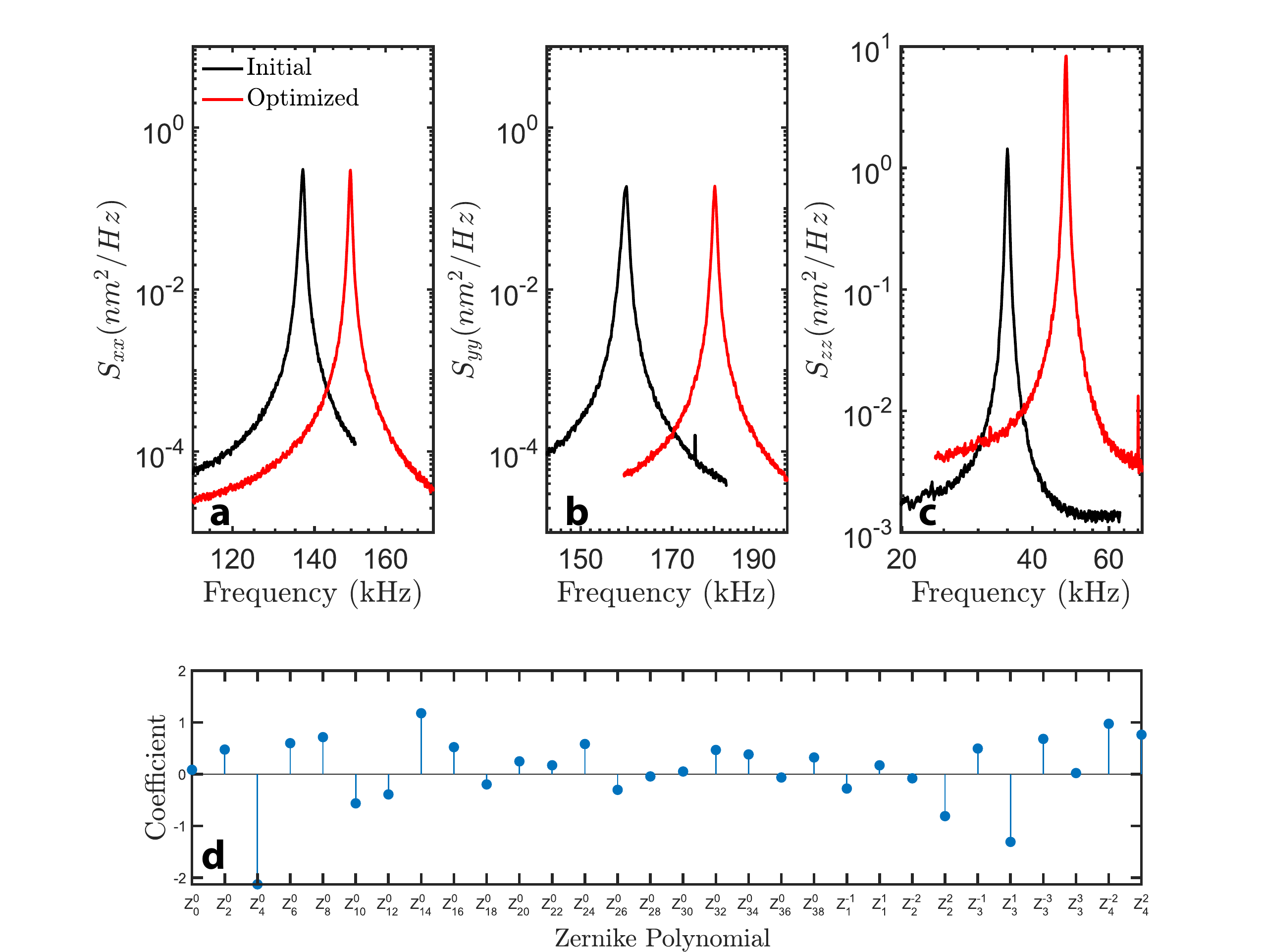}
    \caption{\textbf{Uniform vs optimized spectra}. \textbf{a}, \textbf{b} and \textbf{c}- PSDs for the uniform (black) and optimized (red) wavefront, which are measured along the $x$, $y$ and $z$-axis, respectively.
    \textbf{d}- Relative contributions of the different Zernike polynomials to the optimized wavefront used in \textbf{a}, \textbf{b} and \textbf{c} (see Fig. \ref{fig:zernike_poly}).
    }
    \label{fig:psd}
\end{figure}
To provide insights into how the wavefront is reshaped, Fig. \ref{fig:psd} focuses on the optimization routine being applied to a nanoparticle with a 125 nm radius (which does not correspond to the one used in the main text).
The Figs. \ref{fig:psd}\textbf{a-c} provide the PSDs for each axis measured before ('Uniform', black) and after ('Optimized', red) the optimization. 
Figure \ref{fig:psd}\textbf{d} displays the relative distribution of the different Zernike polynomials (Fig. \ref{fig:zernike_poly}) that compose the optimized wavefront.
It can be observed that the axis-symmetric polynomials (i.e., $Z_i^0$ with $i\in [0,38]$) of lower rank seem to be more involved. 

At last, the robustness of the optimized-wavefront solution is further explored in Fig. \ref{fig:particle_size}.
This figure shows the relative stiffness enhancement for particles of different sizes using the optimized pattern obtained in Figure 1 of the main article. 
These results indicate that a pattern optimized for a given size remains effective for particles of different radii. 
Nonetheless, the optimization is less efficient than when performed directly on the particle of the proper dimension. 
\setcounter{figure}{3}
\begin{figure}[h!]
    \centering
    \includegraphics[width=421pt,height=298pt,trim={3cm 9cm 3cm 10cm},clip]{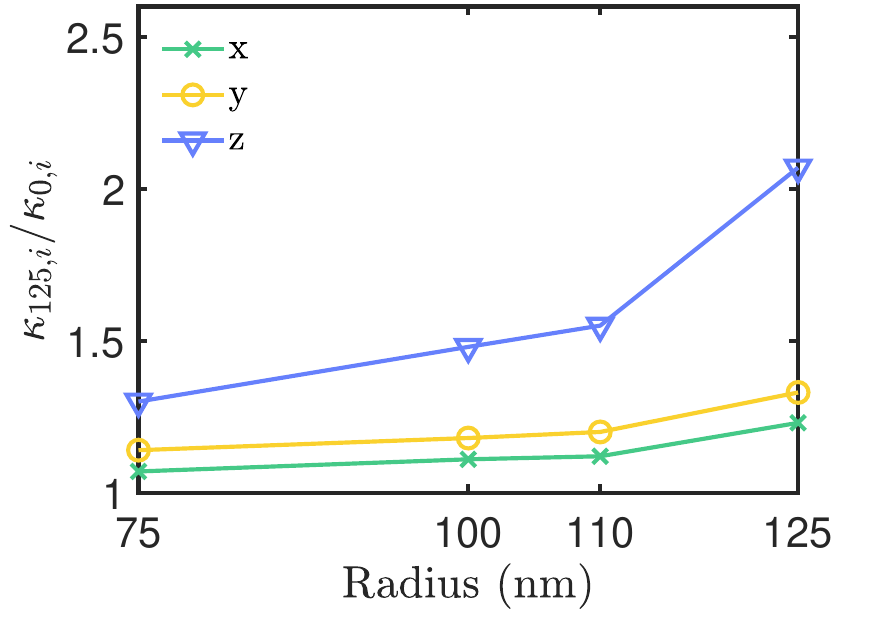}
    \caption{\textbf{Stiffness enhancement ratios obtained for different radii when applying the wavefront from Figure 1 of the main text}. {Relative stiffness} for each axis as a function of the particle's size, using the pattern from the main text. The notation $\kappa_{125}$ indicates that the wavefront was optimized for a particle of 125 nm in radius.}
    \label{fig:particle_size}
\end{figure}

\newpage
\section{Numerical simulations of the optimization process}
\subsection{Forces computation and multipole expansion}
\setcounter{figure}{6}
To model the trapping field for various numerical apertures, NA, filling factors, $f_0$, and wavefronts, $X$, we use a modified Debye integral. 
Specifically, we include a thin-lens apodization function (see equation (3.56) of Ref \cite{novotny2012principles}), which accounts for the SLM-modulated wavefront. 
The field scattered by the particle is expressed using the Generalized Lorentz Mie Theory (GLMT), which ultimately enables to compute 'exactly' the optical forces using the Maxwell Stress Tensor (MST) \cite{Gouesbet2019}. 
This method is used to benchmark a much faster semi-analytical method, which provides the total force as a sum of different multipole contributions \cite{Riccardi2022, Chen2011}. 

\begin{figure}[h!]
    \begin{subfigure}{0.49\textwidth}
        \includegraphics[scale=0.4, trim={-.5cm 7cm .5cm 8cm},clip]{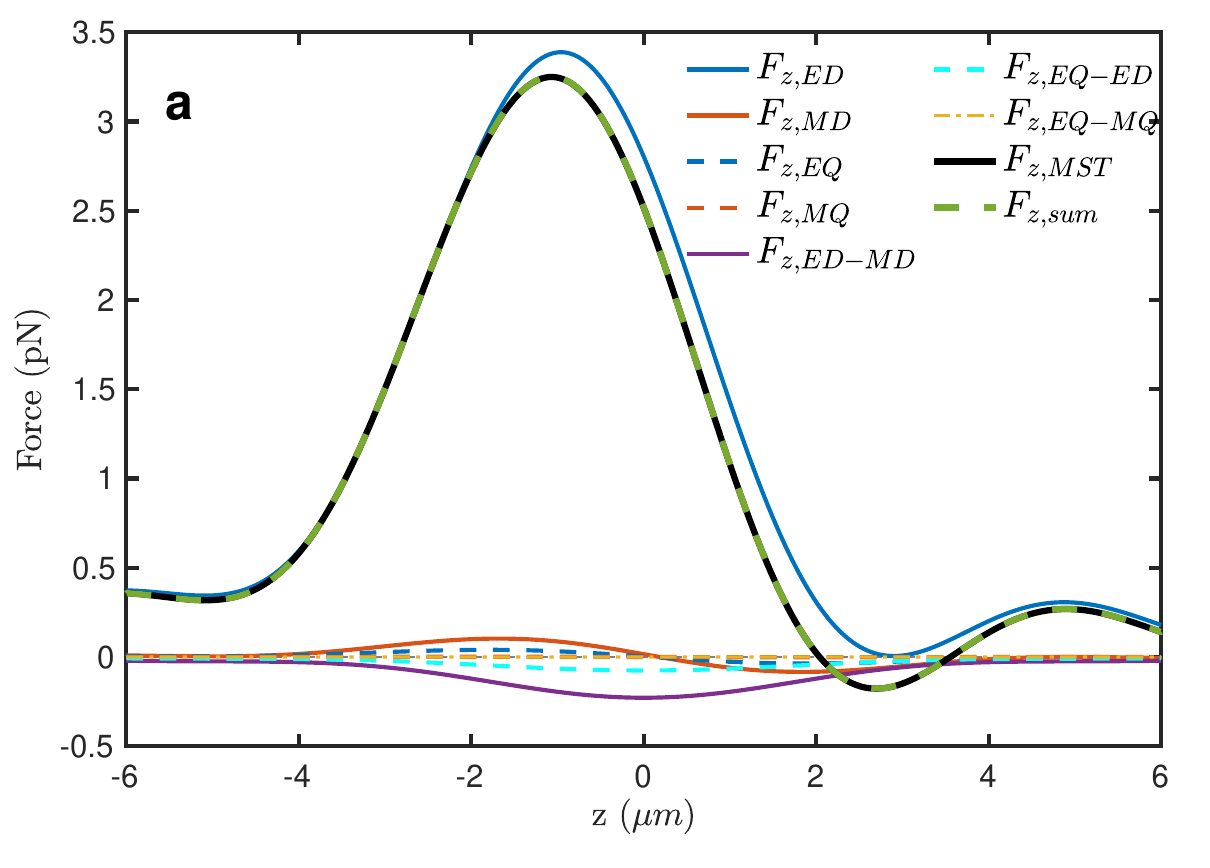}
    \end{subfigure}
    \begin{subfigure}{0.49\textwidth}
        \includegraphics[scale=0.4, trim={-.5cm 7cm .5cm 8cm},clip]{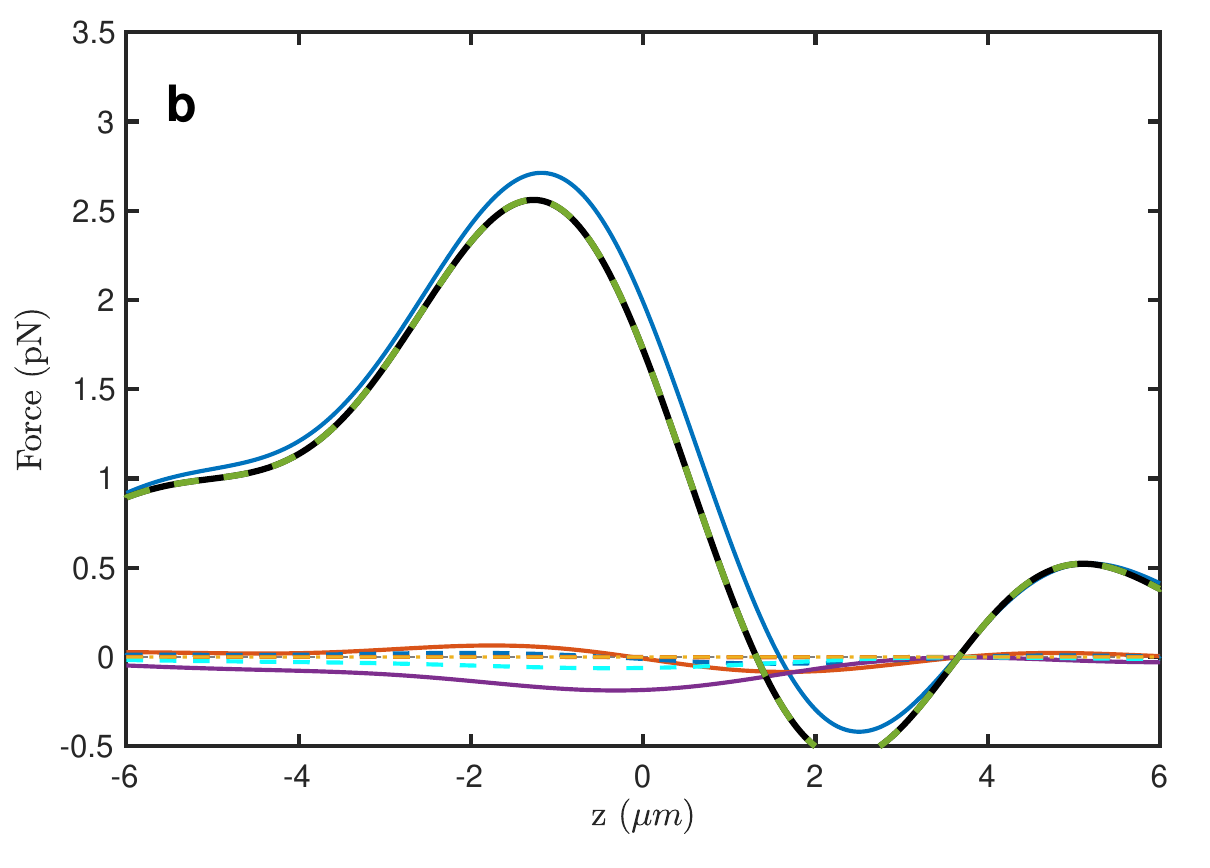}
    \end{subfigure}
\vspace{-2 cm}
    \begin{subfigure}{0.2\textwidth}
        \includegraphics[scale=0.6,trim={1.2cm 0cm 0cm 4cm},clip]
        {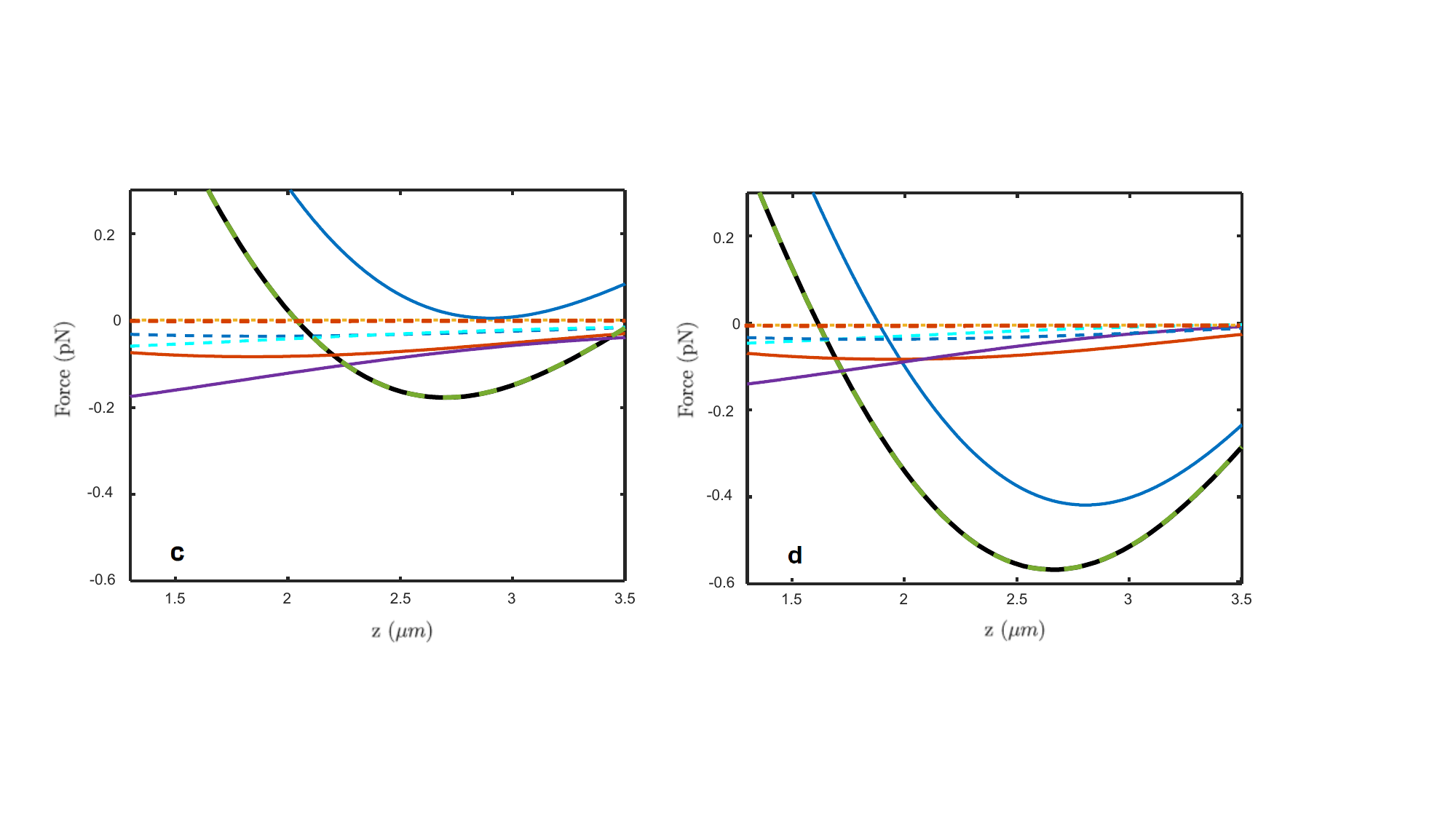}

    \end{subfigure}

    
    \caption{\textbf{Simulations of axial-force landscapes}. \textbf{a}- Landscape of the axial force, $F_z$, for a uniform wavefront. 
    The black curve represents the 'exact' MST computation, $F_{z,MST}$. 
    The other curves (see legend and text) describe the different multipole contributions. 
    When adding up all these contributions, we obtain a total force (dashed green, $F_{z,sum}$) identical to the one computed using the MST. 
    \textbf{b}- Similar force landscape to that in \textbf{a} but here for an optimized wavefront (same legend and color code).
    \textbf{c} and \textbf{d} show magnified views of \textbf{a} and \textbf{b} around their corresponding trapping points, respectively.}
    \label{fig:theo_axial_opt}
\end{figure}
The optimization routine is performed by computing, at each iteration, the force landscapes along $x$, $y$ and $z$, labeled $F_x(x)$, $F_y(y)$ and $F_z(z)$, respectively.
Such computations are achieved using the multipole method (introduced above) up to the quadrupole order.
The modulated wavefront is reproduced by decomposing the apodization function into $32$ to $128$ concentric rings with different optical phases. 
The distribution of the phase on each ring emulates a wavefront $X$, which is iteratively optimized to maximize a cost function similar to the one provided in section \ref{sec:Opt}. 
Varying the number of rings, as well as the starting guess, we observe a convergence towards wavefronts similar to the ones obtained experimentally. 
Figure \ref{fig:theo_axial_opt} shows in black the axial-force landscape, $F_{z,MST}(z)$, computed for a uniform (panel \textbf{a}) and an optimized wavefront (panel \textbf{b}), which display a gain in axial stiffness $\kappa_z$ close to a factor of 2.2 (see Figure 2 of the main text). 
Zoomed-in views of both landscapes are provided in Figs. \ref{fig:theo_axial_opt}\textbf{a} and \textbf{d}.
Here, the simulation parameters are set to an NA=$0.68$ and a filling factor $f_0=0.7$, for a particle of radius 125 nm trapped using a beam power of 350 mW.
We underline, however, that a stiffness gain higher than 2 (as discussed in this work) has been observed numerically for a wide range of particle radii, numerical apertures and filling factors.
We also note that the coordinate $z$ in the optimized case has been shifted such that the conservative part of the force (i.e., gradient force, see section \ref{sec:ConVSNonCon}) remains zero at $z=0$ (i.e., the beam focus).

\setcounter{figure}{8}
\begin{figure}[h!]
    \begin{subfigure}{0.49\textwidth}
        \includegraphics[scale=0.45, trim={1cm 7cm 2cm 8cm},clip]{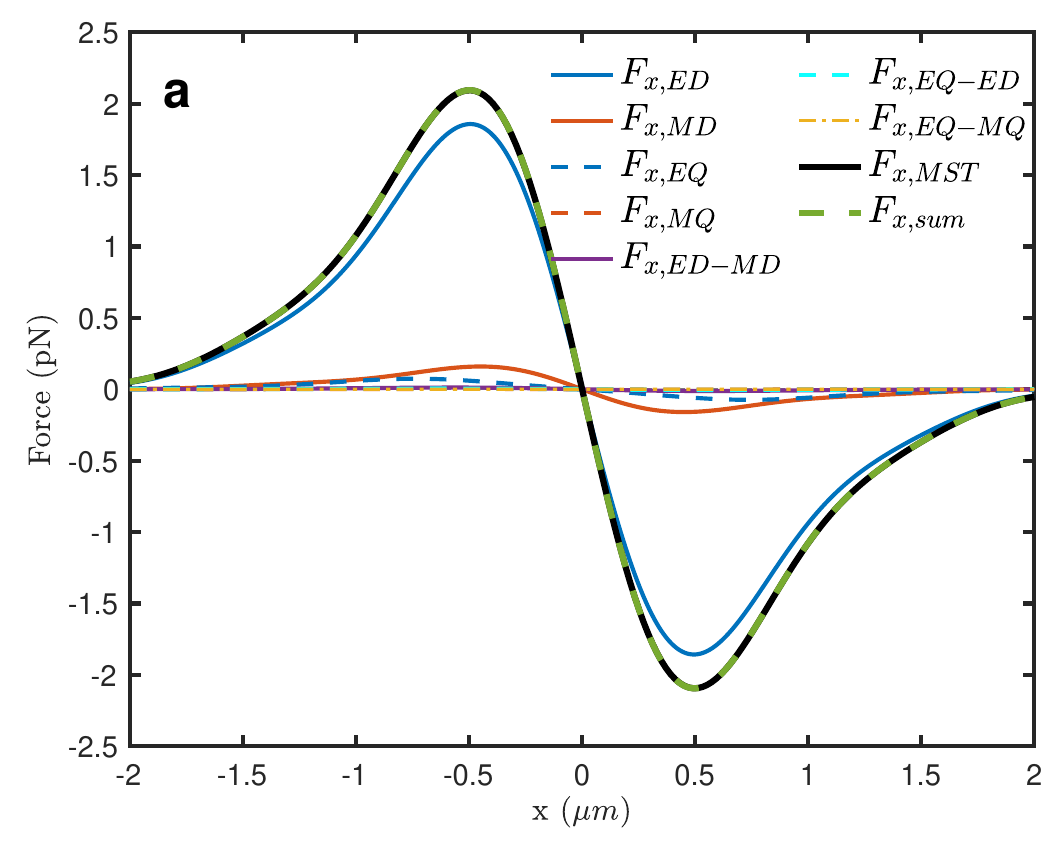}
    \end{subfigure}
    \begin{subfigure}{0.49\textwidth}
        \includegraphics[scale=0.45, trim={1cm 7cm 2cm 8cm},clip]{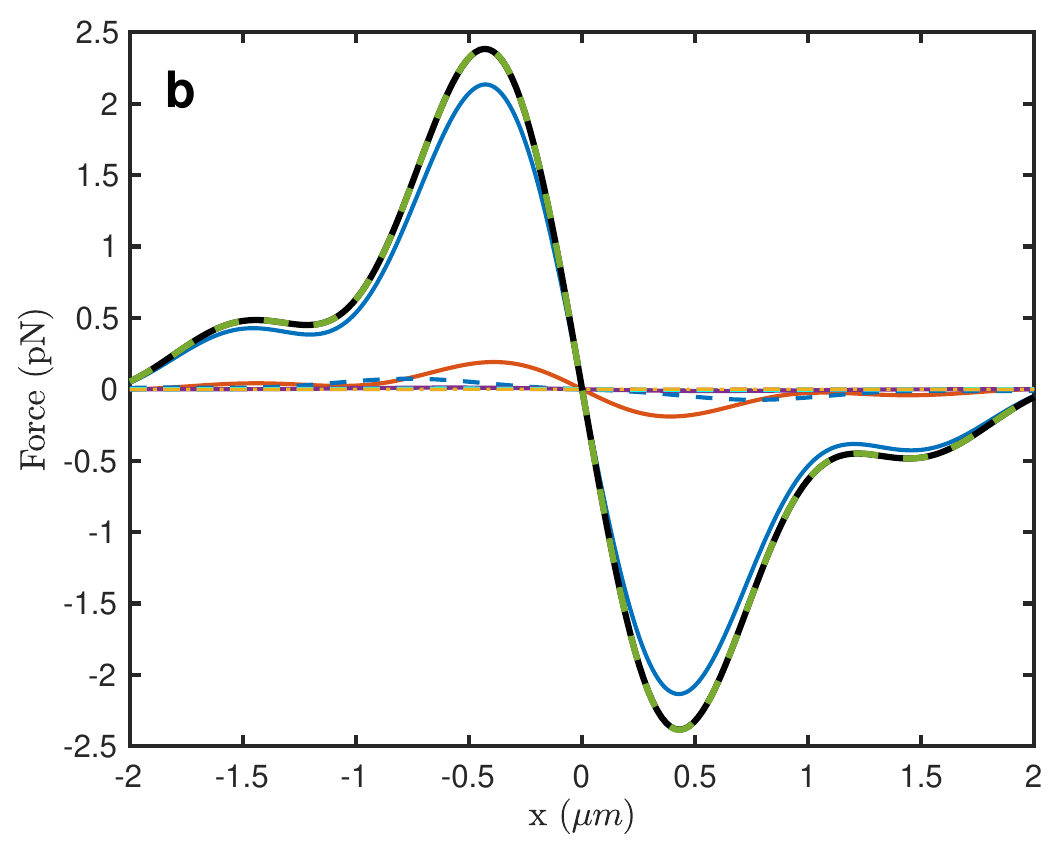}
    \end{subfigure}
    \begin{subfigure}{0.49\textwidth}
        \includegraphics[scale=0.45, trim={1cm 7cm 2cm 8cm},clip]{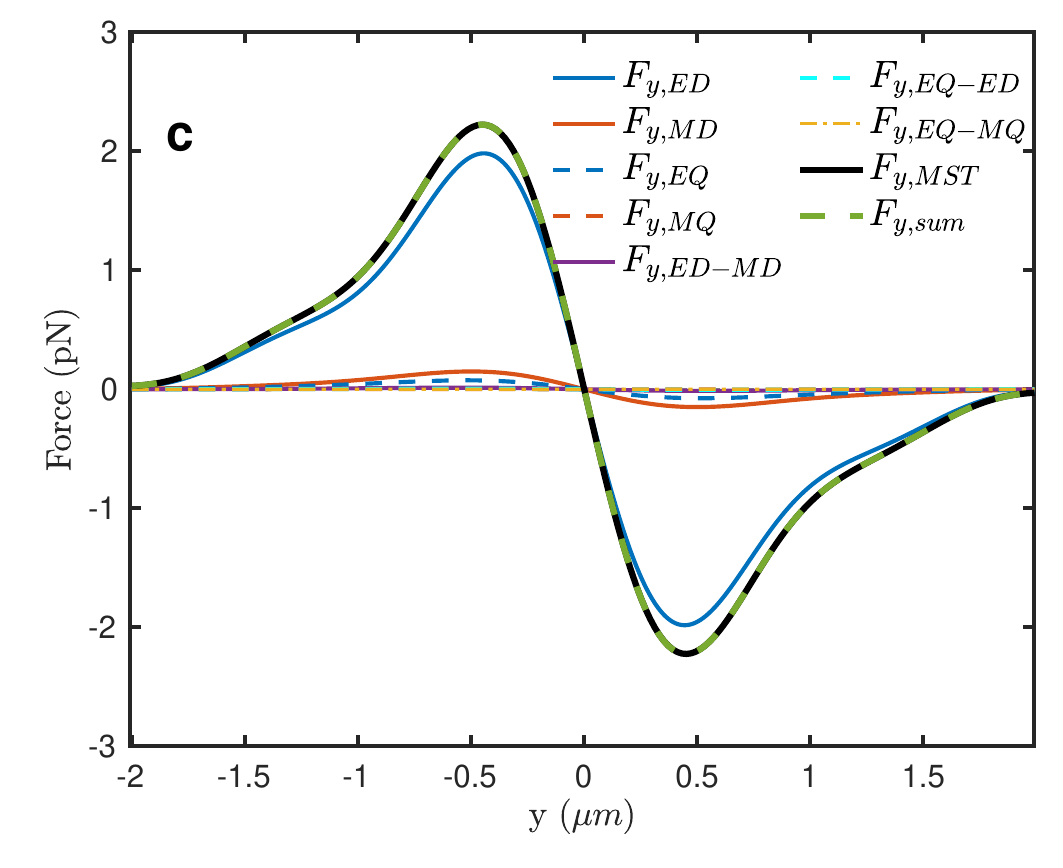}
    \end{subfigure}
    \begin{subfigure}{0.49\textwidth}
        \includegraphics[scale=0.45, trim={1cm 7cm 2cm 8cm},clip]{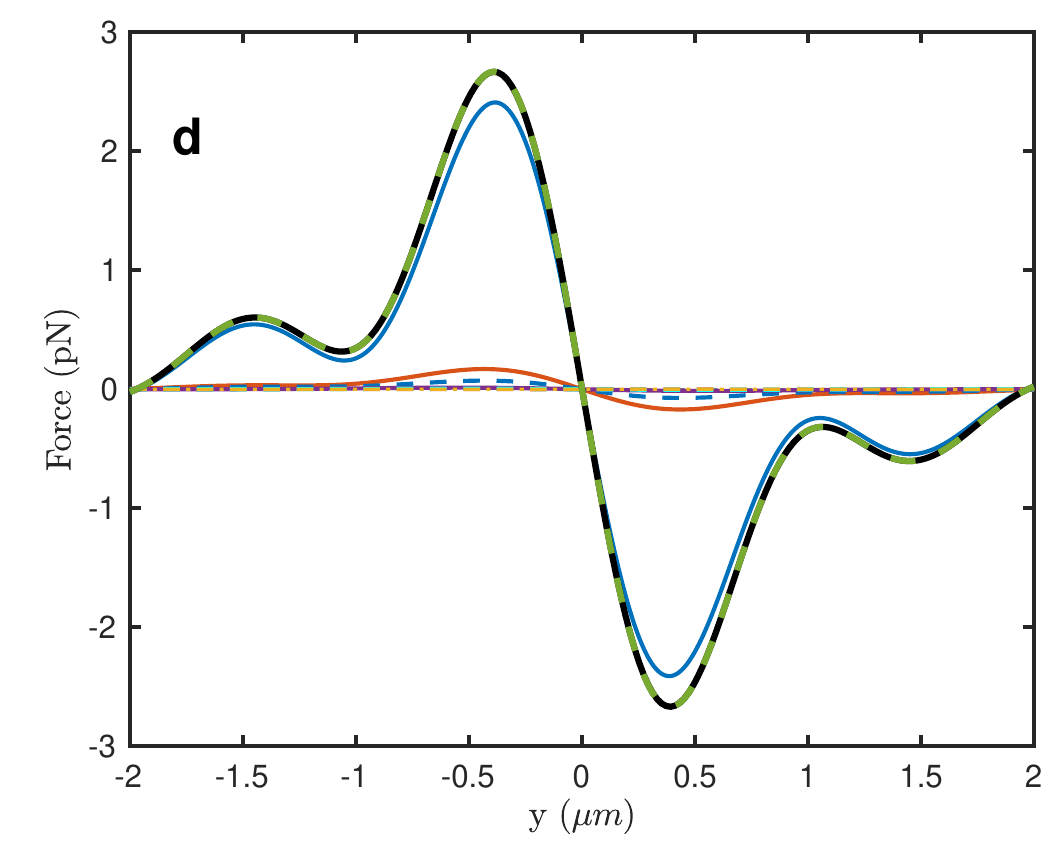}
    \end{subfigure}
    \caption{\textbf{Simulations of transverse-force landscapes}. 
    \textbf{a}, \textbf{b}- Force landscapes $F_x$ computed using the two approaches introduced in Fig. \ref{fig:theo_axial_opt} and for the same wavefronts (respectively uniform in \textbf{a} and optimized in \textbf{b}). 
    \textbf{c}, \textbf{d}- Force landscapes $F_y$ computed using the two approaches introduced in Fig. \ref{fig:theo_axial_opt} and for the same wavefronts (respectively uniform in \textbf{c} and optimized in \textbf{d}). 
    }
    \label{fig:theo_transverse_opt}
\end{figure}
In Figs. \ref{fig:theo_axial_opt}\textbf{a} and \textbf{b}, the different multipole contributions provided by our numerical method are displayed.
In particular, we plot the force originating from the electric dipole (ED, blue), the magnetic dipole (MD, red), the electric quadrupole (EQ, dashed blue) and the magnetic quadrupole (MQ, dashed red).
There, we also report the force produced by the interferences between the electric and magnetic dipoles (ED-MD, purple), between the electric dipole and quadrupole (ED-EQ, dashed light blue) as well as between the electric and magnetic quadrupoles (EQ-MQ, dot-dashed yellow) \cite{Riccardi2022}. 
The curve in dashed green shows the sum of these different multipole contributions, which matches the 'exact' computation performed using the Maxwell Stress Tensor (black). 
In Figs. 2\textbf{a} and \textbf{b} of the main text, the interference terms have been added to the magnetic dipole and electric quadrupole contributions to make these figures easier to understand.
As expected for such a small nanoparticle, these simulations clearly emphasize that the electric dipole is the dominant contribution to the total force along the $z$-axis.
They also show that the optimization mainly acts on the electric-dipole term, while other contributions remain largely unaffected. 
In other words, the optimization primarily reshapes the electric-dipole contribution in order to increase the stiffness.

For the optimization displayed in Figure 2 of the main text and reproduced in Fig. \ref{fig:theo_axial_opt}, Fig. \ref{fig:theo_transverse_opt} shows in black the exact calculations of the forces along the two transverse directions, respectively $F_{x,MST}$ and $F_{y,MST}$.
We also provide the multipole expansion of the forces using the same color code and naming scheme as in Fig. \ref{fig:theo_axial_opt}.
Similarly to the axial direction, we observe that the optimization mainly acts on the dominant electric dipole to improve the stiffness along the transverse directions. 
As sketched in Figure 2c of the main text, the optimization brings the particle closer to the focus, where the intensity gradient is stiffer in the transverse plane, which readily improves the optical confinement.  

\setcounter{figure}{7}
\begin{figure}[h!]
    \begin{subfigure}{0.32\textwidth}
        \includegraphics[scale=0.3, trim={1cm 7cm 2cm 7.5cm},clip]{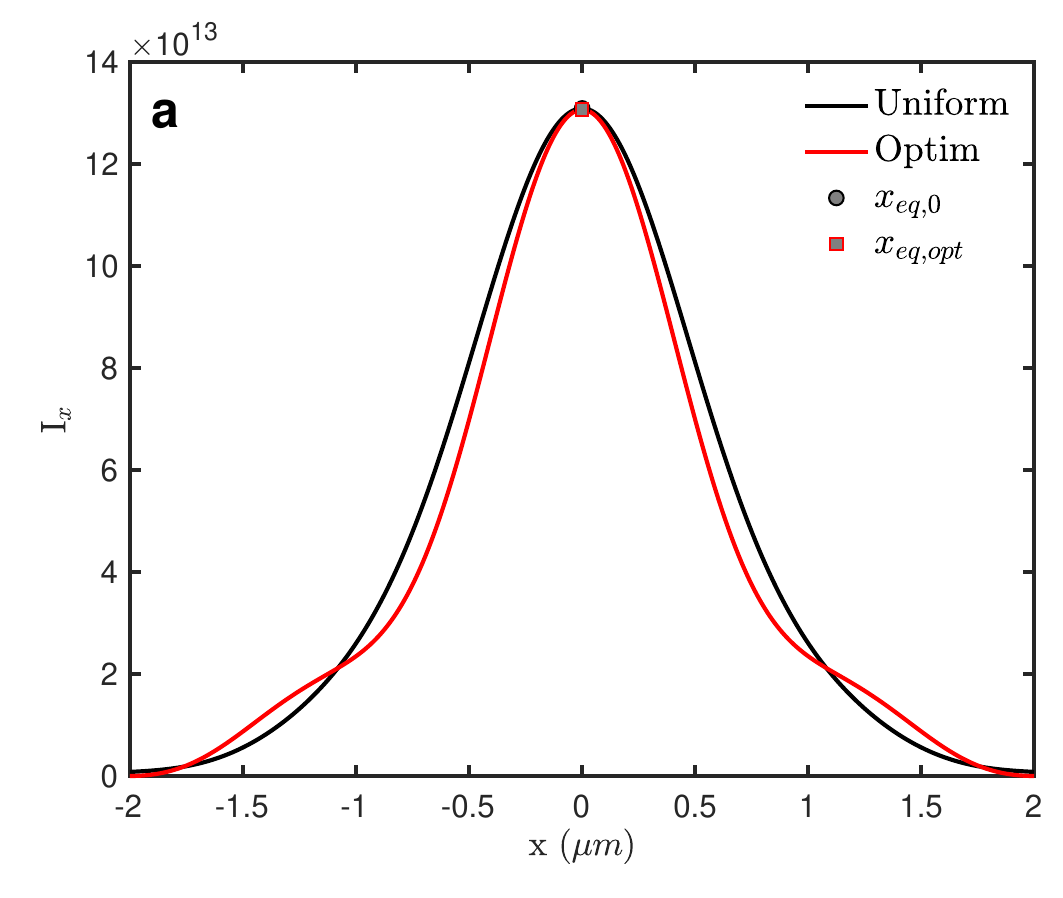}
    \end{subfigure}
    \begin{subfigure}{0.32\textwidth}
        \includegraphics[scale=0.3, trim={1cm 7cm 2cm 7.5cm},clip]{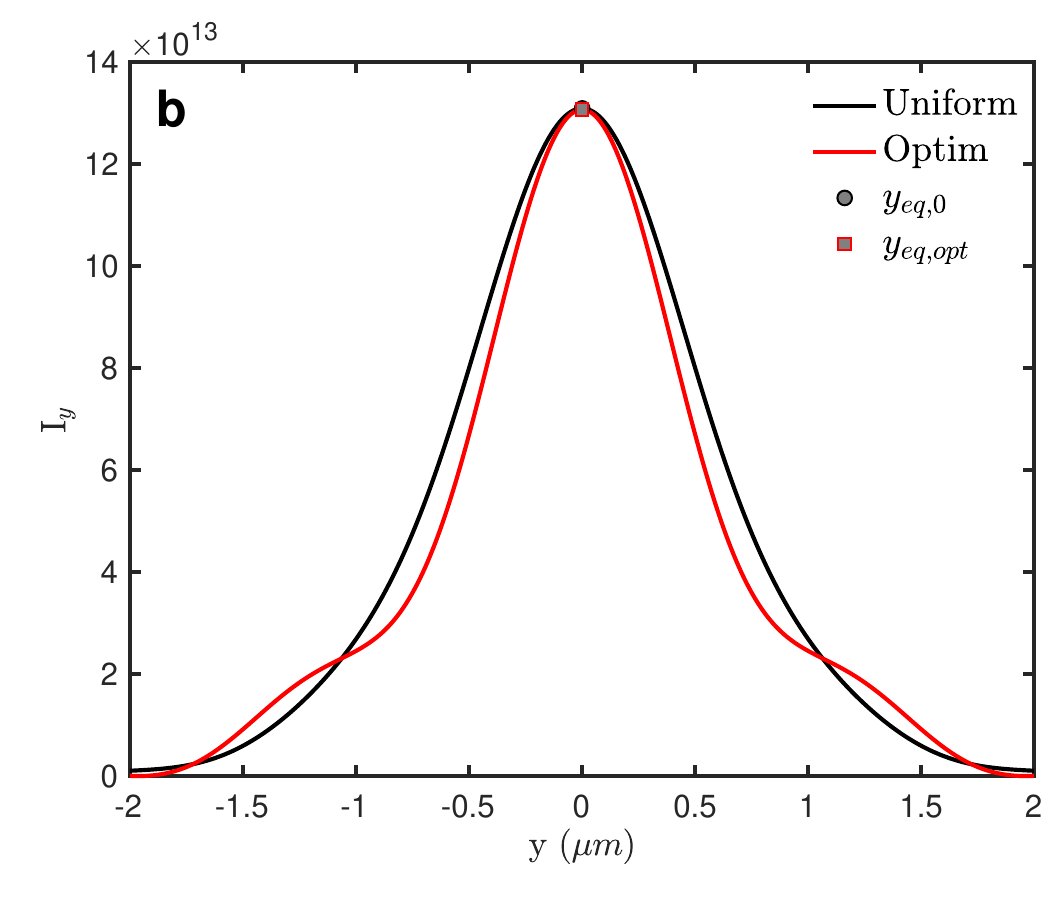}
    \end{subfigure}
    \begin{subfigure}{0.32\textwidth}
        \includegraphics[scale=0.3, trim={1cm 7cm 2cm 7.5cm},clip]{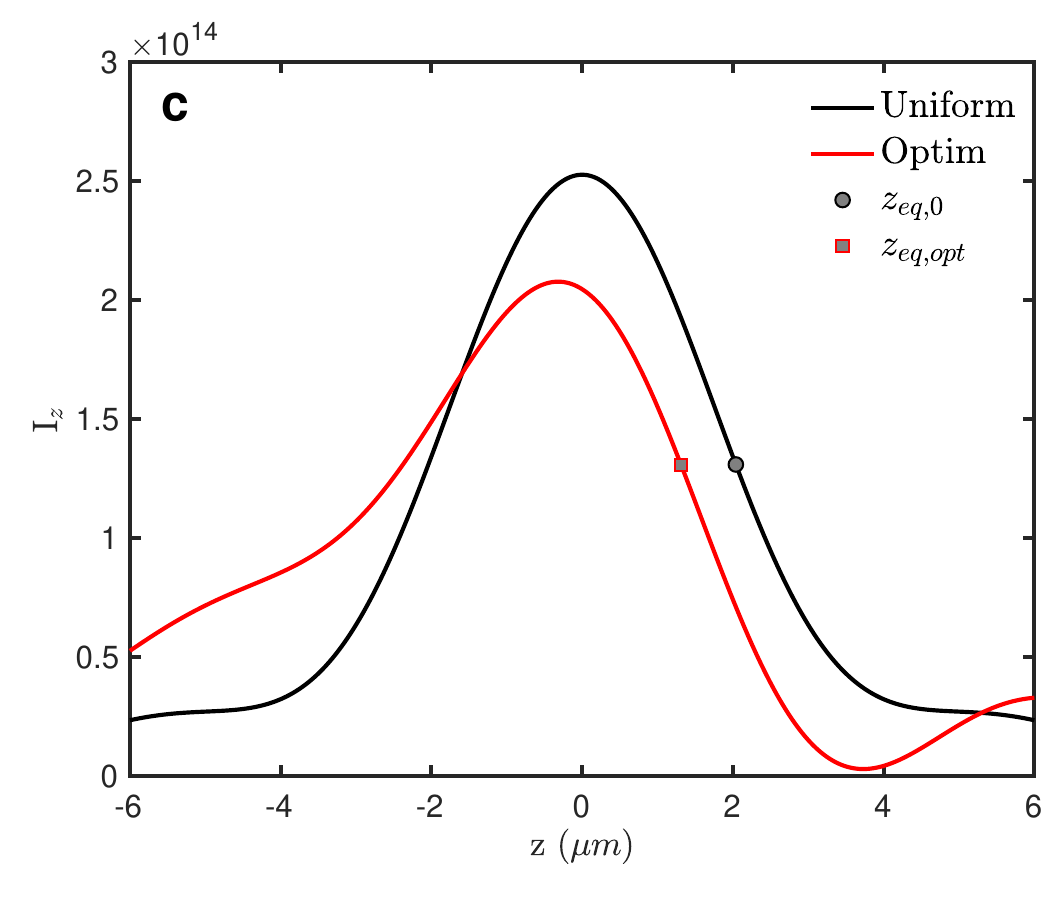}
    \end{subfigure}
    \caption{\textbf{Uniform and optimized local intensities close to the equilibrium positions}. \textbf{a}, \textbf{b} and \textbf{c}- Intensity profiles $I_x$, $I_y$ and $I_z$ (in arbitrary units), which are computed respectively along $x$, $y$ and $z$ for the uniform (red) and optimized (black) wavefronts of Figs. \ref{fig:theo_axial_opt} and \ref{fig:theo_transverse_opt}, respectively. 
    The black squares and red dots pinpoint the equilibrium locations along each axis.}
    \label{fig:theo_intensities}
\end{figure}
After optimization, the optical trap becomes more photon-efficient, in that it can produce the same stiffness as in the uniform case but with significantly less incoming laser intensity.
This point is confirmed in Fig. \ref{fig:theo_intensities}, which plots along the three directions $x$, $y$ and $z$ the field intensities in the vicinity of the focal spot for the uniform (black) and optimized (red) wavefronts.
We readily observe that the intensity at the equilibrium position remains similar in both cases (black squares and red dots for uniform and optimized $z_{eq}$, respectively). 
Thus, as the stiffness is more than doubled by the optimized wavefront (multiplied by $\approx 2.2$), one can achieve the same stiffness as in the uniform case with an incoming laser power reduced by more than $50\%$ (i.e., divided by a factor of 2.2). 

\newpage
\subsection{Conservative and non-conservative parts}
\label{sec:ConVSNonCon}
The multipole expansion performed in Figs. \ref{fig:theo_axial_opt} and \ref{fig:theo_transverse_opt}, can be harnessed to decompose analytically the different terms into their non-conservative (i.e., scattering) and conservative (i.e., gradient) parts \cite{Gouesbet2022}.
Since this decomposition is only provided along $z$ in Fig. 2\textbf{d} of the main text, we only consider the axial forces below.
In particular, the conservative and non-conservative parts are labeled respectively $F_{g,j}$ and $F_{s,j}$, with $j$ indicating the multipole considered.
\setcounter{figure}{9}
\begin{figure}[h!] 
    \centering
        \includegraphics[scale=0.8, trim={0cm 7cm 0cm 7cm},clip]{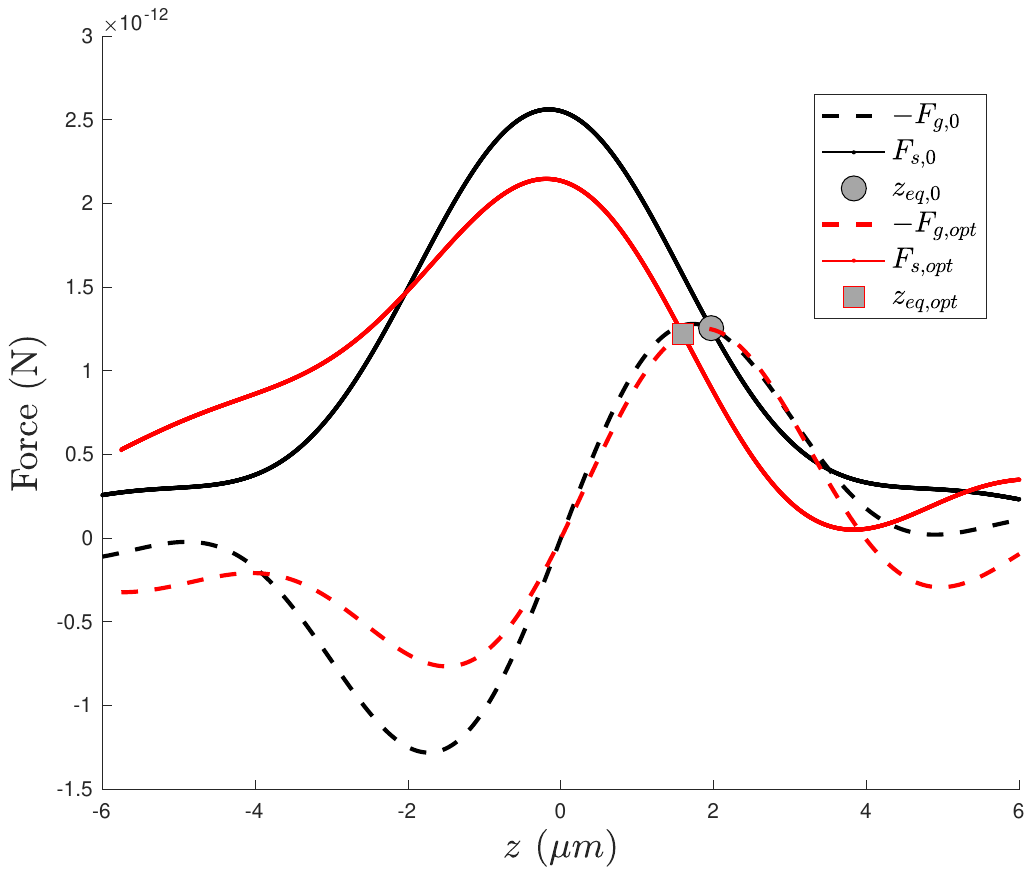}
    \caption{\textbf{Conservative and non-conservative contributions to the axial force}. The total axial force, $F_z$, is decomposed into its conservative and non-conservative parts, see Eqs.~(\ref{F_conserv}, \ref{F_nc}). 
    For the uniform wavefront, we plot in black $-F_{g,0}$ and $F_{s,0}$, which intersect at the equilibrium position $z_{eq,0}$. 
    For the optimized wavefront, we plot in red $-F_{g,opt}$ and $F_{s,opt}$, which intersect at the equilibrium position $z_{eq,opt}$. }
    \label{fig:theorie_GradVSScat}
\end{figure}

For the conservative parts:
\begin{eqnarray}
\label{F_conserv}
F_{g,{ED}}&=& \frac{\varepsilon_0}{2} \rm{Re}\left(\alpha_{ED}\right).\rm{Re}\left(\frac{\partial E_j}{\partial z} E_j^\star \right) \\
F_{g,{MD}}&=& \eta_0^2 \frac{\varepsilon_0}{2} \rm{Re}\left(\alpha_{MD}\right).\rm{Re}\left(\frac{\partial H_j}{\partial z} H_j^\star \right) \nonumber
\label{eq:cons}
\end{eqnarray}

while, for the non-conservative parts:
\begin{eqnarray}
\label{F_nc}
F_{s,{ED}}&=& \frac{\varepsilon_0}{2} \rm{Im}\left(\alpha_{ED}\right).\rm{Im}\left(\frac{\partial E_j}{\partial z} E_j^\star \right) \\
F_{s,{MD}}&=& \eta_0^2 \frac{\varepsilon_0}{2} \rm{Im}\left(\alpha_{MD}\right).\rm{Im}\left(\frac{\partial H_j}{\partial z} H_j^\star \right) \nonumber \\
{F_{s,ED-MD}}&=& - \frac{k^4}{12 \pi \varepsilon_0 c} \rm{Re}\left( \alpha_{ED}~\alpha_{MD}^\star . \left[ \mathbf{E} \wedge \mathbf{H^\star}\right] \cdot \mathbf{u_z} \right),\nonumber
\label{eq:Ncons}
\end{eqnarray}
, where $j \in [x,y,z]$ and summation over repeated indices is implied. Here, 
 $\eta_0$ stands for the vacuum impedance, $k$ for the wavenumber, and $(\mathbf{E}, \mathbf{H})$ refer to the trapping field (uniform or optimized). 
 At last, $\alpha_{ED, MD}$ stand respectively for the complex polarizabilities of the electric and magnetic dipoles, while $\mathbf{u_z}$ corresponds to the unitary vector along $z$.\\

The conservative, $F_g$, and non-conservative part, $F_s$, of the total force $F_z$ are obtained by summing the different expressions provided respectively on Eqs.~(\ref{eq:cons}) and (\ref{eq:Ncons}).
Note that, as the quadrupole terms make only small contributions to the total force in the present case (see Figs. \ref{fig:theo_axial_opt}), we can safely assume that they do not affect the mechanism at play. 
As a result, we can indifferently incorporate their contributions onto either $F_s$ or $F_g$ (here, we chose the former).
Figure \ref{fig:theorie_GradVSScat} displays $-F_g$ (dashed) and $F_s$ (solid) for a uniform (black, subscript $0$) and an optimized wavefront (red, subscript $opt$).
The intersection of both curves defines the equilibrium position, $z_{eq}$. 
We clearly observe that, after the optimization, the conservative force $F_g$ remains almost identical in the vicinity of the equilibrium position.
In sharp contrast, we observe that the non-conservative force $F_s$ is largely shifted towards the focus.

\newpage
\section{Brownian vortices}
\subsection{Theoretical framework}
In liquids, non-conservative scattering forces in optical traps are known to give rise to non-equilibrium probability currents, commonly referred to as Brownian vortices\cite{sun2009brownian}. 
These currents have also been reported for trapped particles governed by underdamped motions, as demonstrated in both theoretical\cite{Yacine_th} and experimental studies\cite{Yacine_exp}. 

We denote $P(\mathbf{x},\mathbf{v},t)$ the probability distribution in position and velocity space ($\mathbf{x}$ and $\mathbf{v}$, respectively) of a nanoparticle over time, $t$. 
The evolution of this probability distribution is governed by a Fokker-Planck equation 
\begin{equation}
    \frac{\partial P(\mathbf{x},\mathbf{v},t)}{\partial t}= -H P(\mathbf{x},\mathbf{v},t)=-\mathbf{\nabla _x}\mathbf{J_x}(\mathbf{x},\mathbf{v},t)-\mathbf{\nabla_v}\mathbf{J_v}(\mathbf{x},\mathbf{v},t)
\end{equation}
, in which $\mathbf{J_x}$ and $\mathbf{J_v}$ stand for the space and velocity probability currents, respectively. 
These currents fulfill
\begin{equation*}
    \begin{split} 
        &\mathbf{J_x} = \mathbf{v}P(\mathbf{x},\mathbf{v},t)\\
        &\mathbf{J_v} = -\frac{k_B T \gamma}{m^2} \mathbf{\nabla_v}P(\mathbf{x},\mathbf{v},t)-\frac{\gamma}{m}P(\mathbf{x},\mathbf{v},t) +\frac{1}{m}\mathbf{F}_t(\mathbf{x})P(\mathbf{x},\mathbf{v},t)
    \end{split} 
\end{equation*}
, where $\gamma$ stands for the friction coefficient, $m$ the particle's mass, $\mathbf{F}_t$ the total force acting on the particle, $k_B$ the Boltzmann constant, and $T$ the temperature.
Derivations based on a minimal scattering model (MSM) with a Gaussian field distribution indicate that the amplitude of Brownian vortices is directly affected by the distribution of the scattering force\cite{Yacine_th}.
Thus, a change in the probability currents is indicative of a change in the scattering-force landscape. 

\subsection{Experimental measurement of probability currents}
Experimentally, the probability $P(\mathbf{x},\mathbf{v},t)$ is estimated using the following expression
\begin{equation}
    P(\mathbf{x},\mathbf{v},t) = \langle\delta(\mathbf{x}-\mathbf{X}_t)\delta(\mathbf{v}-\mathbf{V}_t)\rangle
\end{equation}
, where $\langle . \rangle$ denotes the statistical average, while $\mathbf{X}_t$ and $\mathbf{V}_t$ represent the instantaneous position and velocity of the particle at time $t$, respectively.
The effective probability currents are then given by
\begin{equation}
    \begin{split} 
        &\overline{\mathbf{J_x}} = \langle \mathbf{V}_t\delta(\mathbf{x}-\mathbf{X}_t) \rangle \\
        &\overline{\mathbf{J_v}} = \langle \mathbf{\dot{V}}_t \delta(\mathbf{v}-\mathbf{V}_t) \rangle
    \end{split} 
\end{equation}

These effective currents can be accurately determined in the underdamped regime from temporal traces of the nanoparticle's position using standard conditional binning histograms.
The photodiodes produce electric signals, $V(t)$, which relate to the nanoparticle's motion, $x(t)$, through a calibration factor $C_{calib}$ (in $V/m$) fulfilling in the spectral domain ${S_{VV}}(\Omega)=C_{calib}^2{S_{xx}}(\Omega)$.
This calibration factor is estimated using the equipartition theorem applied to the kinetic energy, $E_{kin}$, of the nanoparticle \cite{article:calib_novotny}
\begin{equation}
    \langle E_{kin}\rangle = \frac{1}{2}m\frac{\langle\dot{V^2}\rangle}{C_{calib}^2}=\frac{1}{2}k_BT
\end{equation}
, where the position variance is given by
\begin{equation}
    \langle x^2\rangle = \int^{\infty}_0 d\Omega {S_{xx}}(\Omega)
\end{equation}

$C_{calib}$ is computed for each wavefront along the three axes, assuming a $T=$ 300K temperature at high pressures ($> 1$ mbar).
Temporal traces are filtered around the resonance frequency of each axis, yielding the particle's position relative to its equilibrium position. 
Velocities and accelerations are then calculated using a Gaussian kernel function, which minimizes noise by applying a locally weighted average to the data.
This calibration method ensures accurate conversion of raw sensor data into real physical quantities, enabling reliable measurements of velocities and accelerations even in the presence of experimental noise.

\subsection{Experimental results}
\begin{figure}[h!] 
    \centering
    \includegraphics[scale=0.4]{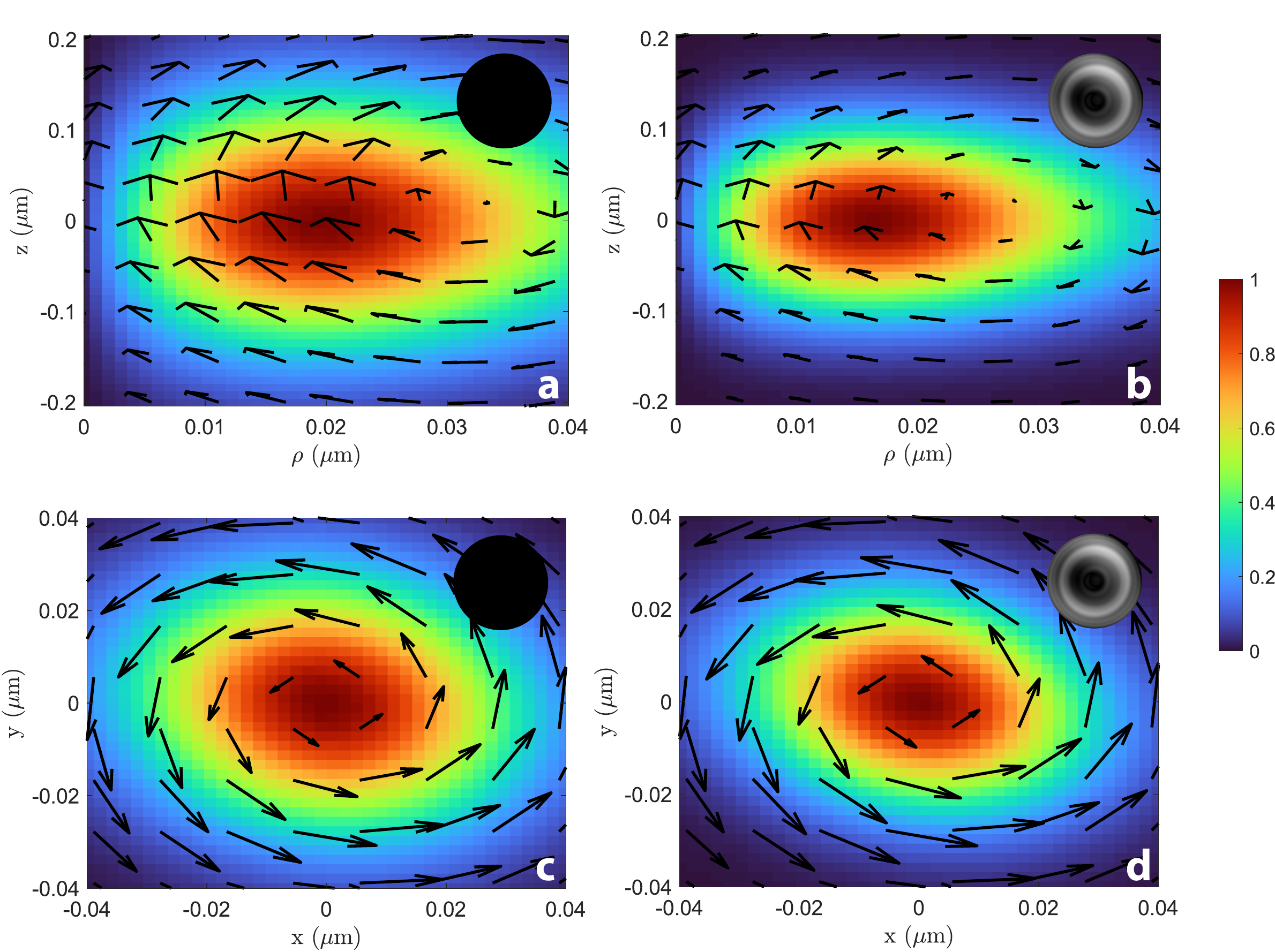}
    \caption{\textbf{Scattering force-induced vortices in position space}. Probability distributions $P_x$ (colored map) and currents $\overline{J_x}$ (black arrows) in the space $(\rho , z)$ for a 125 nm radius particle trapped at 1 mbar usinf a uniform (\textbf{a}) and an optimized wavefront (\textbf{b}).
    Probability distributions $P_x$ (colored map) and currents $\overline{J_x}$ (black arrows) in the space $(x , y)$ for a 125 nm radius particle trapped at 1 mbar using a uniform (\textbf{c}) and an optimized wavefront (\textbf{d}).
    The amplitudes of the distributions $P_x$ are normalized to their maximum values. }
    \label{fig:rouleaux_position}
\end{figure}

\begin{figure}[h!] 
    \centering
        \includegraphics[scale=0.45]{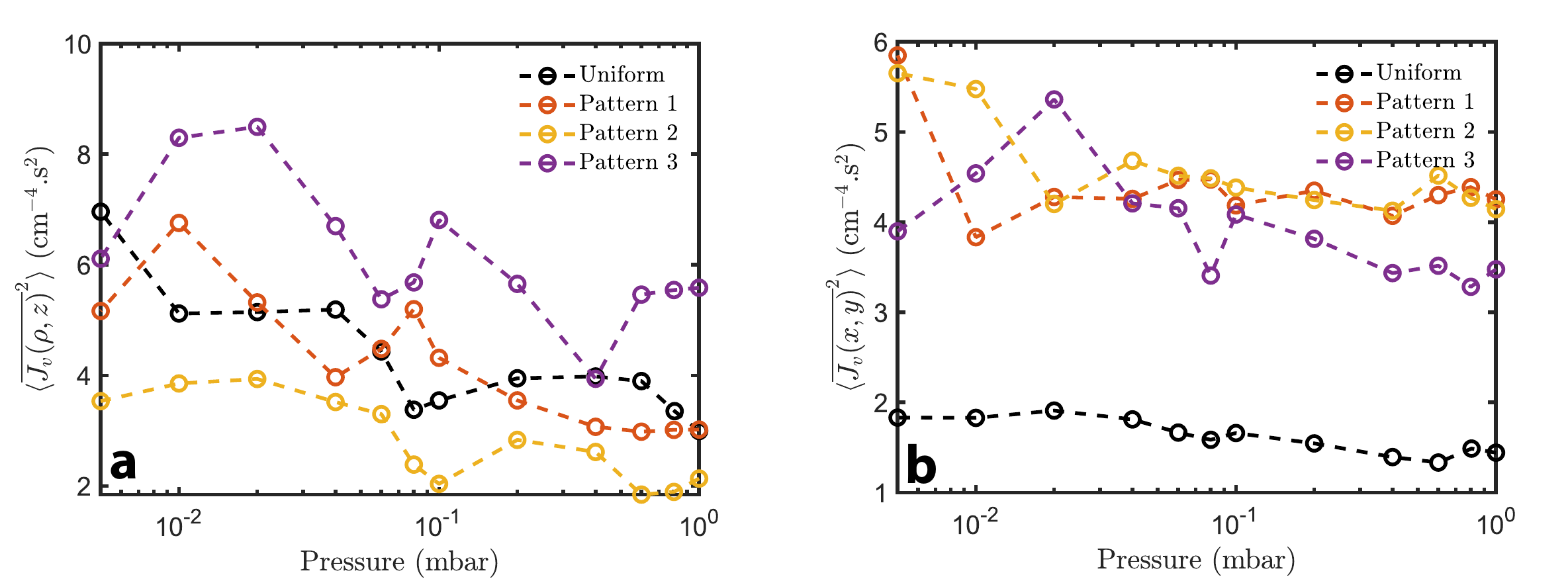}
    \caption{\textbf{Vortices in velocity space}. Amplitude of the vortices as a function of pressure for a 110 nm radius particle, using the wavefronts obtained from the 3 optimizations provided in Fig. \ref{fig:iter}. 
    Panel \textbf{a} considers the space $(v_{\rho}, v_z)$, while panel \textbf{b} considers the space $(v_x , v_y)$.}
    \label{fig:Jv_pressure}
\end{figure}

Figures \ref{fig:rouleaux_position}\textbf{a} and \textbf{b} illustrate, for respectively a uniform and an optimized wavefront (see insets), the probability currents in the position space $(\rho, z)$, where $\rho = \sqrt{x^2 + y^2}$ represents the transverse axis in cylindrical coordinates. 
We clearly observe that the currents display vortices, whose amplitude is altered when the optimized wavefront is applied. 
Figures \ref{fig:rouleaux_position}\textbf{c} and \textbf{d} display, for respectively a uniform and optimized wavefront, the probability currents in the position space $(x, y)$. 
These results demonstrate more pronounced confinements of the particle's distributions in all spatial directions when using the optimized wavefront (i.e., indicating enhanced trapping stiffnesses). 
Additionally, the amplitude of the Brownian vortices is significantly altered, particularly in the $(\rho, z)$ space, reflecting the impact of wavefront optimization onto scattering forces.
Figure \ref{fig:Jv_pressure} compares the vortex amplitudes measured in velocity space for the three phase patterns obtained from the optimizations in Fig. \ref{fig:iter}, when pressure is reduced. 
Although the amplitude increases consistently in the space $(v_x , v_y)$ compared to the uniform wavefront, this trend is not observed in the space $(v_{\rho}, v_z)$. 
These results suggest that wavefront shaping can be leveraged to modulate optical scattering effects.

\newpage
\section{Reducing nonlinearities}
The resonance frequency of an optically trapped nanoparticle is expected to remain stable in a harmonic potential. 
However, as the particle explores larger oscillation amplitudes, nonlinear effects introduce deviations in the resonance frequency.
Those nonlinear frequency shifts become particularly significant in the underdamped regime, where low damping leads the system beyond the linear response region.

\begin{figure}[h!]
    \centering
    \includegraphics[scale=0.35]{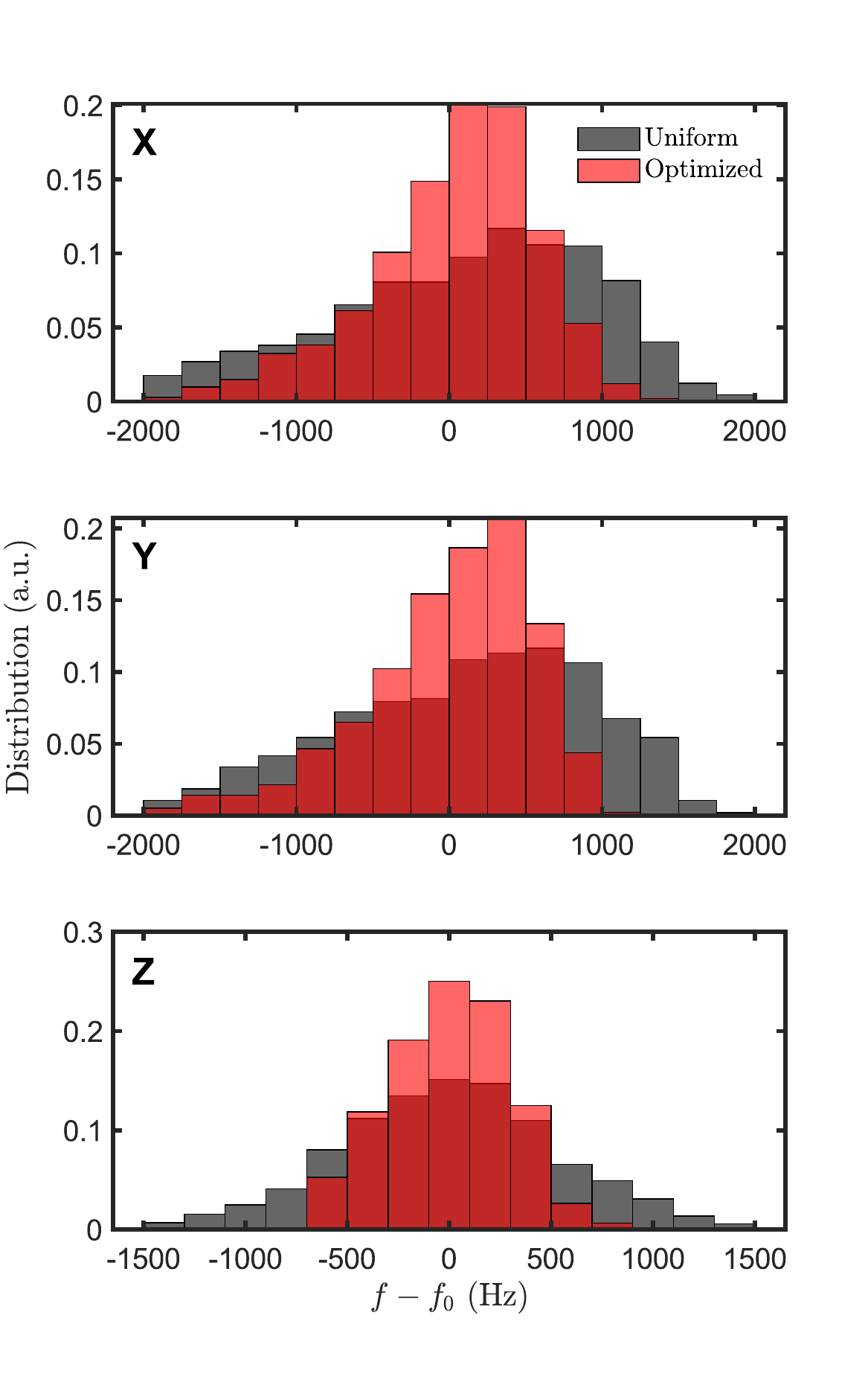}
    \includegraphics[scale=0.35]{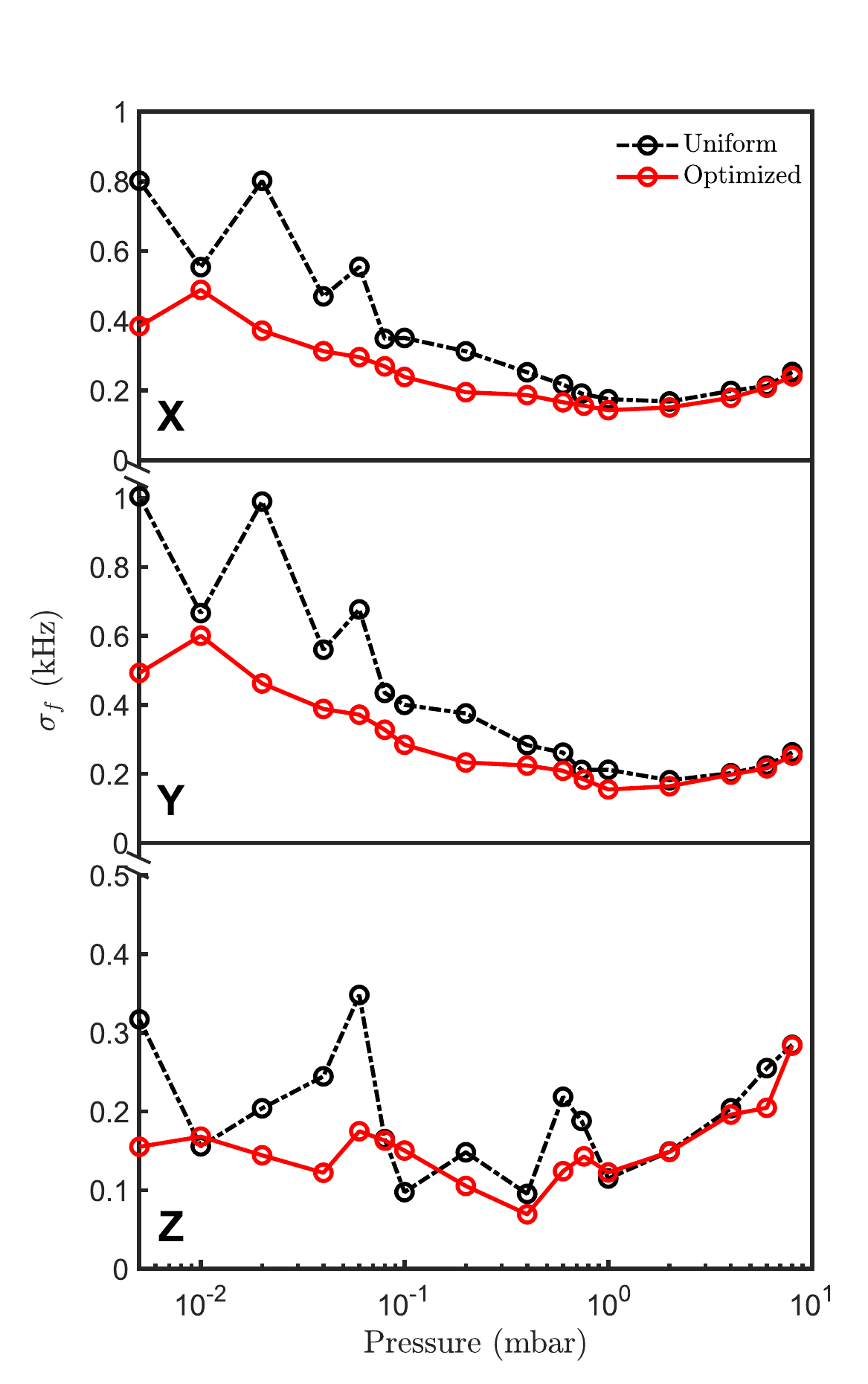}
    \caption{\textbf{Nonlinearities}. \textbf{a} Histograms of the resonance frequency distribution at 0.1 mbar for the three axes, comparing a uniform (black) and an optimized wavefront (red). 
    \textbf{b} Standard deviations, $\sigma_f$, of 10 ms time-trace frequency distributions measured along $x$, $y$ and $z$ at different pressures for a uniform (black) and an optimized (red) wavefront.}
    \label{fig:hist}
\end{figure}

To characterize the nonlinear effects displayed in Figure 4 of the main text, we analyze the frequency fluctuations as a function of pressure. 
Following the approach of Ref \cite{Gieseler2013}, the resonance frequency is extracted from short-time PSDs computed over 10 ms intervals. 
The statistical distribution of resonance frequencies was then determined from a 20 s total trace, yielding a dataset of 2000 traces per pressure value.
Figure \ref{fig:hist}\textbf{a} shows histograms of resonance frequencies for a uniform (black) and an optimized (red) wavefront along each axis at 0.1 mbar (using the same nanoparticle studied in the main article). 
On Figure 4\textbf{c} of the main text, we plot the standard deviation $\sigma_f$ of these distributions as a function of pressure. 
As a consistency check regarding the reduction of nonlinearities, we reproduce in Fig. \ref{fig:hist}\textbf{b} the same approach using a different 125 nm particle.

The purpose of the modulation was to increase the trap stiffness, thereby enhancing the confinement of the particle. 
Our results show that this increased confinement delays the onset of nonlinearities as the pressure is reduced (Fig. \ref{fig:hist}\textbf{b}). 
In the uniform case, nonlinear frequency fluctuations appear at higher pressures, whereas in the optimized configuration, the system remains in the linear regime over a broader pressure range. 
This demonstrates that wavefront shaping not only enhances trap stiffness but also provides a means to control and mitigate nonlinear effects.

\newpage

\bibliographystyle{unsrt}

\renewcommand{\refname}{Sources} 
\bibliography{biblio_MP} 
\end{document}